\documentclass[12pt,preprint,number,sort&compress,lontitle,times]{elsarticle}
\usepackage{natbib} 
\usepackage{amsmath} 
\usepackage{amssymb}
\usepackage{xspace}
\usepackage{graphicx}
\usepackage{epsfig}
\usepackage{lineno}
\usepackage{multirow}
\usepackage{fourier}
\usepackage{color}
\usepackage{hyperref}
\usepackage{cleveref}
\usepackage{appendix}
\usepackage[english]{babel}
\usepackage[]{geometry}
\newcommand{\dif}{\mathrm{d}}
\newcommand{\LDF}{\ensuremath{{\rm LDF}}\xspace}
\newcommand{\rref}{\ensuremath{r_{\rm ref}}\xspace}
\newcommand{\sref}{\ensuremath{S_{\rm ref}}\xspace}
\newcommand{\lgsref}{\ensuremath{\lg S_{\rm ref}}\xspace}
\newcommand{\rhot}{\ensuremath{R_{\rm hot}}\xspace}
\newcommand{\ssat}{\ensuremath{S_{\rm sat.}}\xspace}

\begin{document}

\begin{frontmatter}

\title{An Improved Method of Estimating the Uncertainty of Air-Shower Size at Ultra-High Energies}
\author[1]{A. Coleman\corref{cor}}
\ead{alanco@umich.edu}
\author[2]{P. Billoir}
\ead{billoir@lpnhe.in2p3.fr}
\author[3]{O. Deligny}
\ead{deligny@ijclab.in2p3.fr}

\cortext[cor]{Corresponding author}

\address[1]{University of Delaware, Department of Physics and Astronomy, Bartol Research Institute, Newark, DE, USA}
\address[2]{Laboratoire de Physique Nucl\'eaire et de Hautes Energies (LPNHE), Sorbonne Universit\'e, Universit\'e de Paris, CNRS-IN2P3, Paris, France}
\address[3]{Laboratoire de Physique des 2 Infinis Ir\`ene Joliot-Curie (IJCLab), CNRS/IN2P3, Universit\'{e} Paris-Saclay, Orsay, France}

\begin{abstract}
The collection of a statistically significant number detected of cosmic rays with energy above $10^{17}$ to $10^{18}$\,eV requires widely-spaced particle detectors at the ground level to detect the extensive air showers induced in the atmosphere. 
The air-shower sizes, proxies of the primary energies, are then estimated by fitting the observed signals to a functional form for expectations so as to interpolate the signal at a reference distance.
The functional form describes the rapid falloff of the expected signal with the distance from the shower core, using typically two logarithmic slopes to account for the short-range and long-range decreases of signals.
The uncertainties associated to the air-shower sizes are determined under the assumption of a quadratic dependence of the log-likelihood on the fitted parameters around the minimum, so that a meaningful variance-covariance matrix is provided.
In this paper, we show that for an event topology where one signal is much larger than the others, the quadratic dependence of the fitted function around the minimum is a poor approximation that leads to an inaccurate estimate of the uncertainties.
To restore a quadratic shape, we propose to use the polar coordinates around the detector recording the largest signal, projected onto the plane of the shower front, to define the likelihood function in terms of logarithmic polar distances, polar angles and logarithmic shower sizes as free parameters.
We show that a meaningful variance-covariance matrix is then recovered in the new coordinate system, as the dependence of the fitted function on the modified parameters is properly approximated by a quadratic function.
The use of the uncertainties in the new coordinate system for subsequent high-level analyses is illustrated.
\end{abstract}

\end{frontmatter}
 

\section{Introduction}

Ultra-high energy cosmic rays with energies above $10^{17}$ to $10^{18}$\,eV are observed through the extensive air showers they induce in the atmosphere. 
To collect a statistically significant number of events, very-widely-spaced particle detectors at the ground level have been used to maximize the aperture of the observatories. 
Typical detection methods, such as scintillator panels or water-Cherenkov tanks, observe an integrated charge that results from a convolution of the detector response with the local particle density, energy distribution, and arrival direction. 
The spacing of the particle detectors turns out to be five to ten times larger than the Moli\`ere radius, which delineates the distance in which more than 90\% of the ionizing particles are contained. 
Consequently, it has been an increasingly difficult task to determine the lateral distribution of particles on an event-by-event basis, and thus to use the associated total number of particles 
for estimating the energy.

As an alternative energy estimator, the signal at a distance appropriate to the spacing of the shower array is used as a surrogate measurement of the shower size.
First proposed by Hillas~\cite{Hillas:1970,Hillas:1971}, the technique, successfully exploited by the Haverah Park Collaboration~\cite{Lawrence:1991cc}, was reviewed in details~\cite{Newton:2006wy} and adopted by the Pierre Auger Collaboration, the Telescope Array Collaboration, and others. 
The conversion from shower size into primary energy may subsequently include several correction factors, calibrations, and/or comparisons with simulations. 
The key point of the technique is to infer the shower size from the amount of signal that is expected to be measured by a detector at a reference distance, $\rref$, by fitting a lateral distribution function, \LDF, to the observed data. 
At that specific distance, which depends on the topology of the surface array, the fluctuations in the age of the showers when reaching the ground, inherited from the stochastic variations in the location and character of the leading interactions, reflected in the fluctuations of the \LDF, are minimised. 
In this way, the shower size is determined with adequate accuracy ($<10\%$). 
While a further description of this technique is beyond the scope of this paper, it is important to note that if using a shower-size technique, the resolution with which the size can be determined ultimately sets the minimum resolution on the energy.

This estimation is done in practice by fitting the observed signal amplitudes, $S(r_i)$, at a distance, $r_i$, using an \LDF.
A common convention is to normalize the \LDF using the reference distance, \rref, such that $\LDF(\rref) \equiv 1$ and thus the shower size at the reference distance can be directly extracted from $S(r) = \sref\,\,\LDF(r)$.
The fit of such a model to determine the size of the air shower involves the determination of \emph{at least} five parameters, two which describe the orientation of the shower axis, two which define the intersection of the axis with a plane\footnote{Typically this plane is taken to be the ground in the local coordinate system.}, and \sref.
Additional parameters which describe the exact nature of the exponential decrease in signal size with distance from the shower axis may also be fit, but near the triggering threshold, there may not be enough degrees-of-freedom and an average \LDF is typically used, which is based on observed or simulated air showers. 
The combination of the nearly power-law shape of the \LDF and the first-order cylindrical symmetry of the signals about the shower axis results in a non-trivial phase space for the log-likelihood function (LLH) that can hamper an accurate determination of the optimal parameters as well as their uncertainties.
The aim of this work is to explore carefully the LLH phase space for typical event topologies encountered in practice with contemporary observatories, i.e. the Pierre Auger Observatory and the Telescope Array.

We note that while the arrival direction of the air shower is important for e.g., anisotropy studies, the two parameters that define the intersection point, the \emph{impact point}, are generally nuisance parameters.
This allows for some freedom regarding the choice of parameters to describe the location of the impact point.
In this work, we show how the estimation of the uncertainties, particularly that of the logarithm of \sref, can be determined in a much more stable way when employing a ``log-polar'' coordinate system, rather than a Cartesian one.
In~\cref{sec:simulations}, we set the stage of the generic framework allowing us to develop an efficient tool to simulate and reconstruct events recorded by surface detectors. 
Using this tool, we are able to extract event topologies leading to LLHs that cannot be approximated by quadratic dependencies in the reconstructed parameters.
To correct this issue, we propose in~\cref{sec:polar_coords} to change the coordinate system for the fit, switching to a log-polar one, and show that quadratic dependencies of the LLHs in the reconstructed parameters more accurately describe the true uncertainties. 
Among a variety of possible high-level analyses, we show in~\cref{sec:application} how the energy calibration performed in a manner similar to that used at the Auger Observatory~\cite{Dembinski:2015wqa,PierreAuger:2020qqz} may benefit from accurate event-by-event uncertainties.
Finally, conclusions are given in~\cref{sec:conclusion}.

\section{Simulation and reconstruction of air showers}
\label{sec:simulations}

\begin{figure}[htbp]
    \centering 
    \includegraphics[width=.47\textwidth]{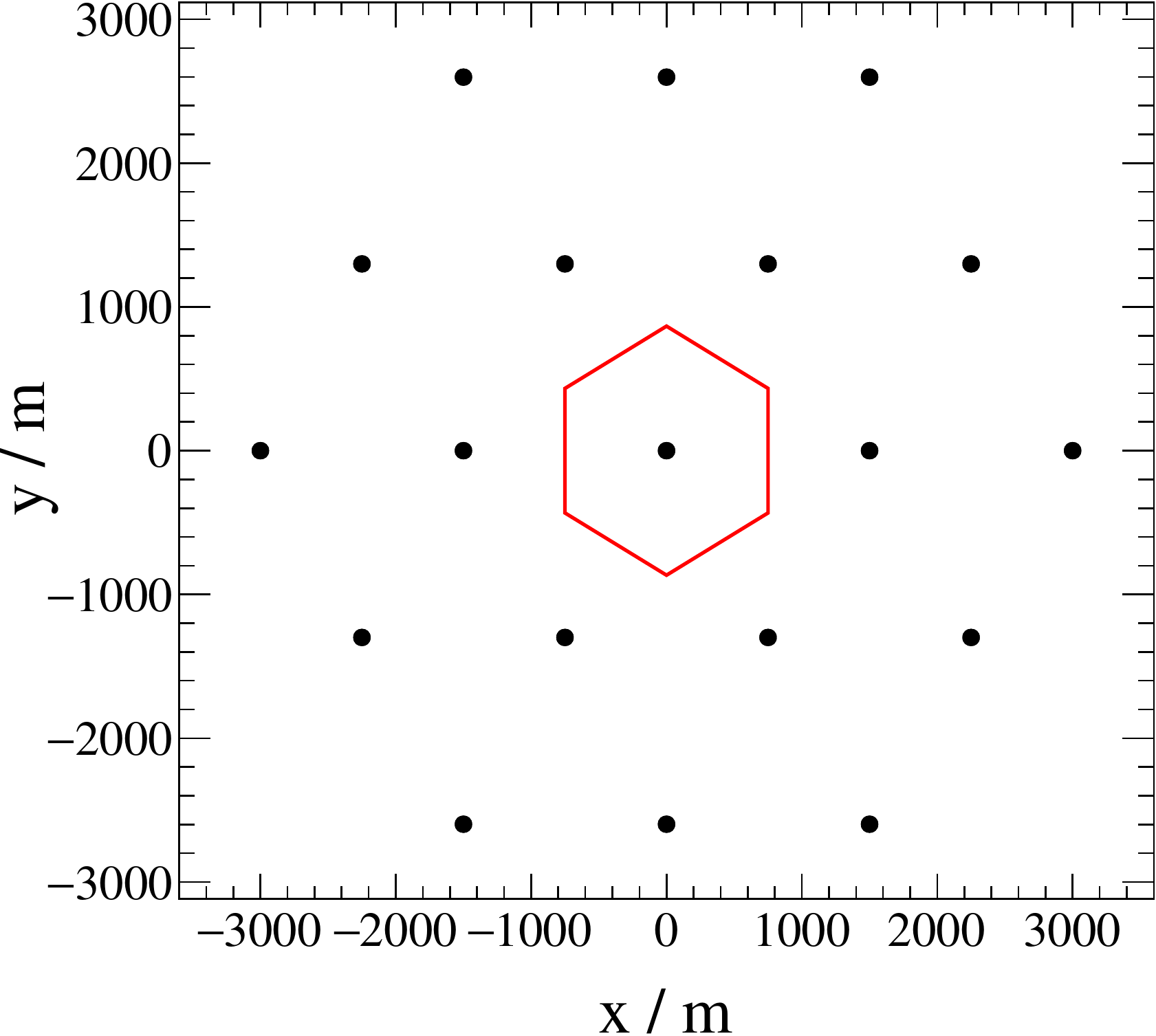}
    \hfill
    \includegraphics[width=.47\textwidth]{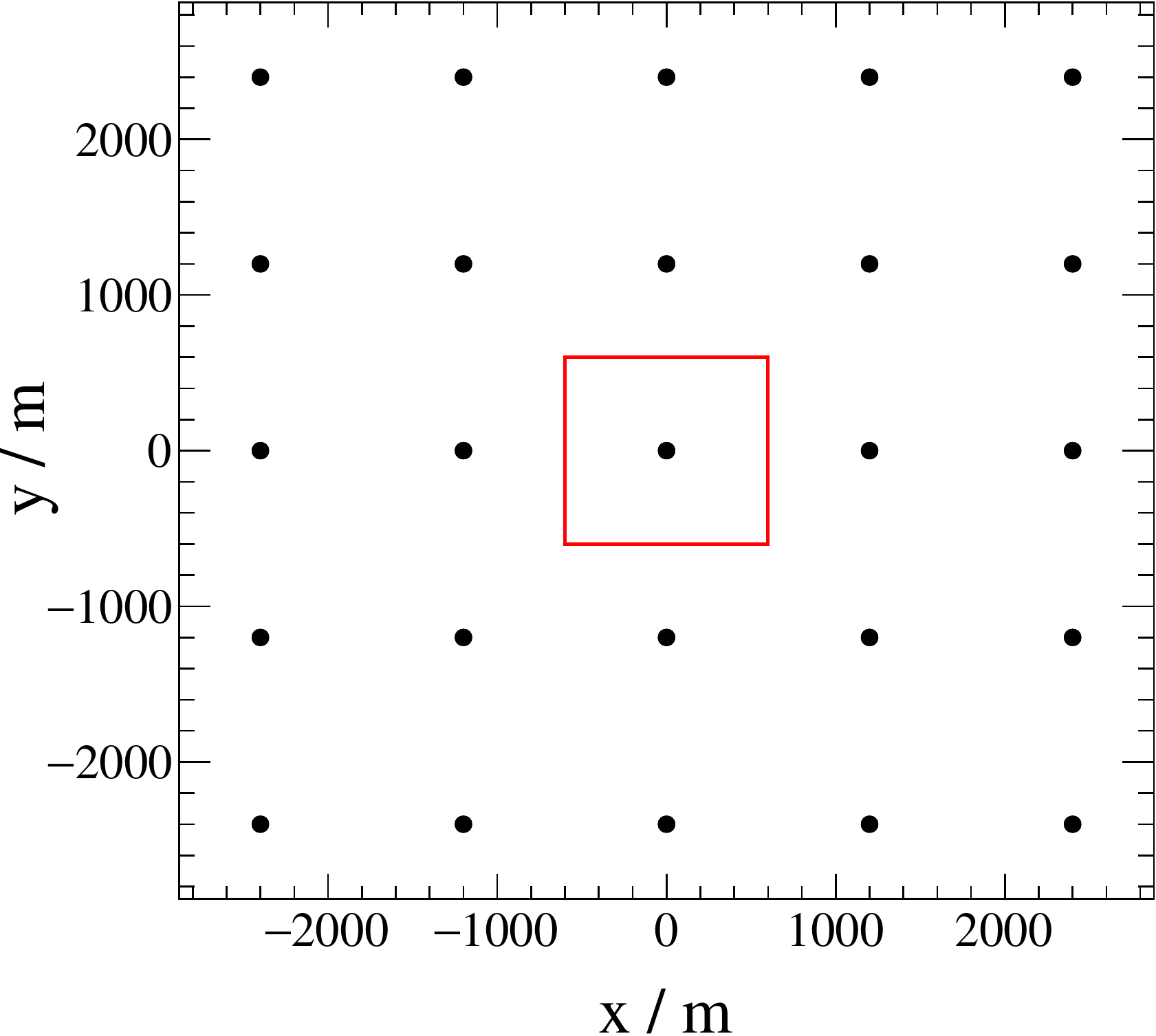}
    \caption{\label{fig:array} 
        The layout of the triangular and square arrays (black circles) are shown in the left and right panels, respectively. The region over which the impact points are sampled for each layout is outlined in red.
    }
\end{figure}

We employ a toy model for the simulation of signals in an air-shower array.
Rather than running e.g., full simulations of the particle cascades, we take the ideal case where the \LDF is known.
Shower-to-shower fluctuations are ignored, as they are nonessential for the purpose of this study.
In this way, the \LDF accurately describes the distribution of signals around the shower axis.
Showers are randomly sampled on a triangular array with 1.5\,km spacing, based on the layout of the Pierre Auger Observatory~\cite{PierreAuger:2015eyc}, see the left panel of \cref{fig:array}.
However, we will also show that the effect is the same for a square array with 1200\,m spacing, based on the arrangement of the surface array of Telescope Array~\cite{TelescopeArray:2012uws}.

Without loss of generality, all signals in this work are expressed in Vertical Equivalent Muon (VEM) units, regardless of the detector or of the particle species crossing the detector.
A VEM is defined as the sum of the charge collected for a single muon traversing vertically a detector (see e.g. ref.~\cite{PierreAuger:2005znw}). 
The value of \sref is sampled randomly from a $\dif N/\dif \sref \propto \sref^{-1}$ spectrum and over a range that is commensurate with energies that correspond to $E > 3$~EeV for the 1500\,m surface detector of the Pierre Auger Observatory.
In this work, we will focus on air showers with a zenith angle of $40^\circ$.
A study on the zenith dependence is shown in the appendix and while a bias on the shower-size estimation is shown to exist at all angles, there is some dependence on the arrival direction in the exact characteristics.

A signal is assigned to each detector on the array using the \LDF,
\begin{equation}
    S(r) = \sref \left( \frac{r}{\rref} \right)^\beta \left( \frac{r + r_0}{\rref + r_0} \right)^{\beta + \gamma},
    \label{eq:ldf}
\end{equation}
with $\beta = -2.5$, $\gamma = 0.1$, $\rref = 1000$\,m, and $r_0 = 700$\,m.
The falloff of the signal from the shower axis is thus described by a logarithmic slope denoted as $\beta$, and by a small departure from a power-law at large distances governed by $\gamma$.
The signal assigned to each detector is additionally smeared using Gaussian fluctuations of width
\begin{equation}
    \sigma(S) = A_S \sqrt{S}.
\end{equation}
The scaling in $\sqrt{S}$ has been shown to describe the Pierre Auger Observatory data~\cite{Ave:2007zz}.
The factor $A_S$, chosen to be 1 in this work, is not considering any dependence in zenith angle in this study\footnote{While a given detector may have a zenith-dependent $\sigma(S)$, here we only study a single zenith angle.}.
Finally, an ideal trigger is implemented which keeps only detectors with a signal above $S_{\rm T} = 3$.
While the values and formulas used above are based on those used for water-Cherenkov tanks at the Pierre Auger Observatory~\cite{PierreAuger:2020yab}, they are similar in nature to those used by other current air-shower experiments~\cite{Newton:2006wy,Yoshida:1994jf,TelescopeArray:2014nxa,IceCube:2012nn,Apel:2010zz} and the issue being discussed here is ultimately a geometric one and is not sensitive to these choices.
Also note that in this work, reconstructions will be performed with a fixed \LDF shape (i.e. $\beta$ and $\gamma$) and thus the choice of \rref is arbitrary and has no influence on the outcome of this study.

\subsection{Reconstructing the air-shower parameters}
\label{sec:reco}

In this study, we mostly study the shower-size estimation.
Since air-shower arrays can typically be used to reconstruct air showers to an accuracy of better than a few degrees~\cite{PierreAuger:2020yab,TelescopeArray:2014nxa,IceCube:2012nn,Apel:2010zz}, we will neglect the reconstruction of the shower arrival direction since it is a sub-dominant effect when reconstructing the shower size and is determined from arrival times rather than signal amplitudes.
Thus in this work, MINUIT is given only three parameters, \lgsref and two parameters that determine the impact point.
In the beginning of this work, the impact point will be defined by the $(x,y)$ position in the ground plane, but we will define a better set of coordinates in \cref{sec:polar_coords}.

The model of an air shower, with the  \LDF shape, is applied to the simulated data and the air shower parameters are fit by minimizing a negative log-likelihood that describes the probability to observe the set of signals, $\{S_i\}$, in each detector.
The likelihood is made of two main terms, one that describes the detectors which triggered, and one for those that did not\footnote{Throughout the paper, the notation $\lg$ stands for the decimal logarithm, while $\ln$ stands for the natural logarithm.},
\begin{equation}
    \label{eq:llh}
    -2\, {\rm LLH} = \sum_{i}^{\rm Triggered} \left[ \left( \frac{S_i - S(r_i)}{\sigma(S(r_i))} \right)^2  + \ln(2 \pi\,\sigma^2(S(r_i))) \right] - 2 \sum_{j}^{S_j = 0} \ln \left( P(S(r_j)) \right).
\end{equation}

The second sum describes the probability, $P(S)$, that no signal was observed when a signal, $S(r_j)$, is expected.
For Gaussian fluctuations, this probability is given by,
\begin{equation}
    P(S) = 1 - \frac{1}{2} \left[ 1 + {\rm erf} \left( \frac{S - S_{\rm T}}{\sqrt{2}\,\, \sigma(S)} \right) \right].
\end{equation}

\subsection{Minimization and error estimation}
\label{sec:minimization}

The reconstruction of an air shower consists of two tasks, traversing through the LLH-space to find the (global, in the ideal case) minimum of \cref{eq:llh} and to characterize the curvature near this minimum to estimate the uncertainty.
Ideally, the whole LLH-space could be scanned and the global minimum can be directly identified.
However, this is typically impractical for a large data set and instead most algorithms require a well-estimated starting point and numerically calculate and follow the gradient of the LLH-space until a minimum is found.
This so-called \emph{gradient-descent} method does not ensure a global minimum is ultimately found and there are various other pitfalls, but a discussion of these issues is beyond the scope of this paper.
Many implementations of this algorithm exist and extensions on this basic concept have been developed, but in this work, we focus on MINUIT~\cite{James:1975dr}, the standard minimizer that is packaged with ROOT~\cite{Brun:1997pa}, which is commonly used in particle and air-shower physics.

To do the second task, MINUIT provides an estimate of the curvature in the neighborhood of the identified minimum in LLH-space.
The Hessian matrix of second-derivatives with respect to all of the free-parameters (impact point, \sref, etc.) is numerically calculated and then inverted to produce the covariance matrix.
This matrix is an estimate of the local curvature assuming that the LLH-space is quadratic near minimum and also provides the correlations between parameters.
This simplistic definition will be shown to be problematic for the reconstruction of a specific class of air showers where the LLH-space does not fulfill the quadratic assumption. 

\begin{figure}[tbp]
\centering 
\includegraphics[width=.49\textwidth]{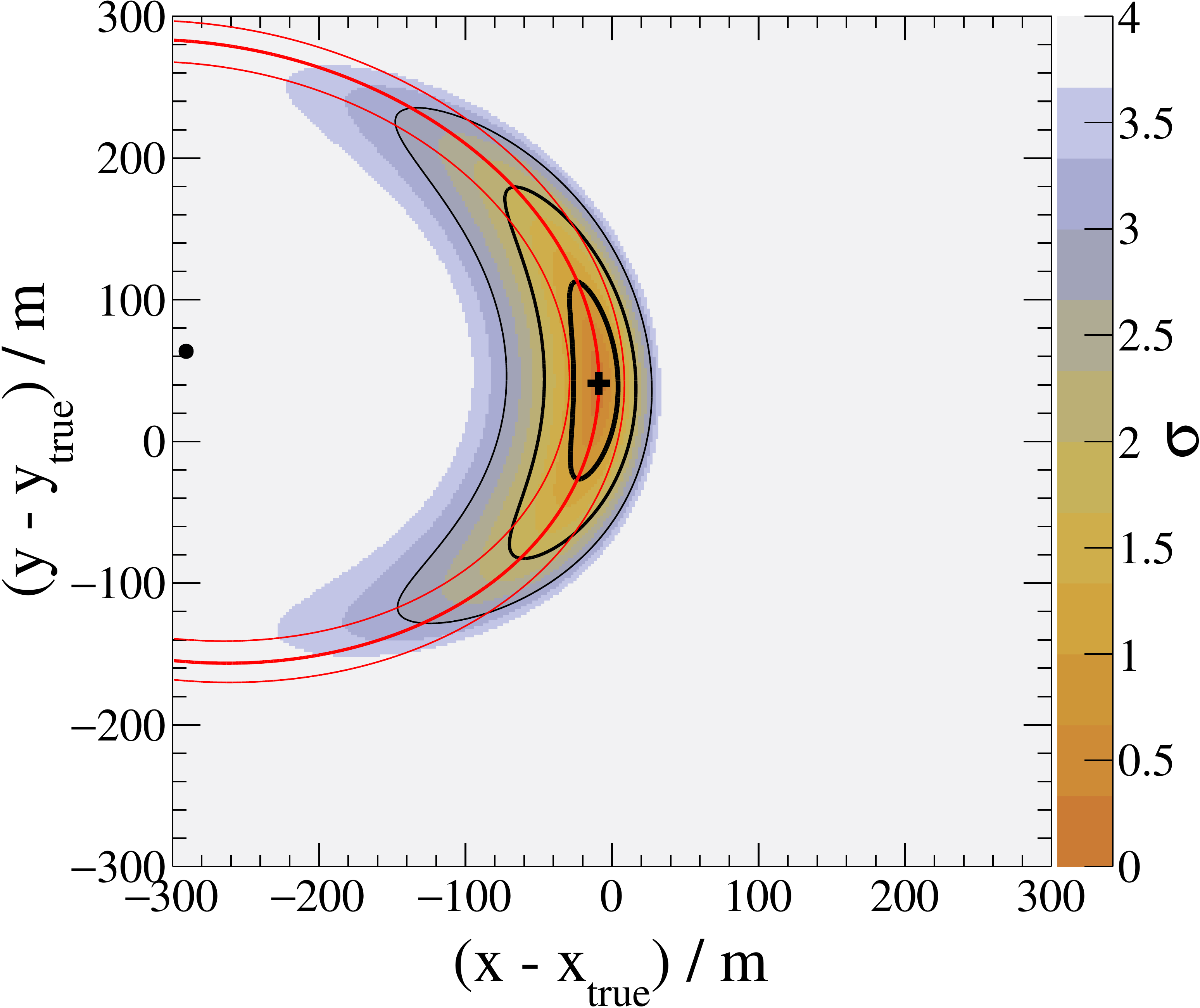}
\hfill
\includegraphics[width=.49\textwidth]{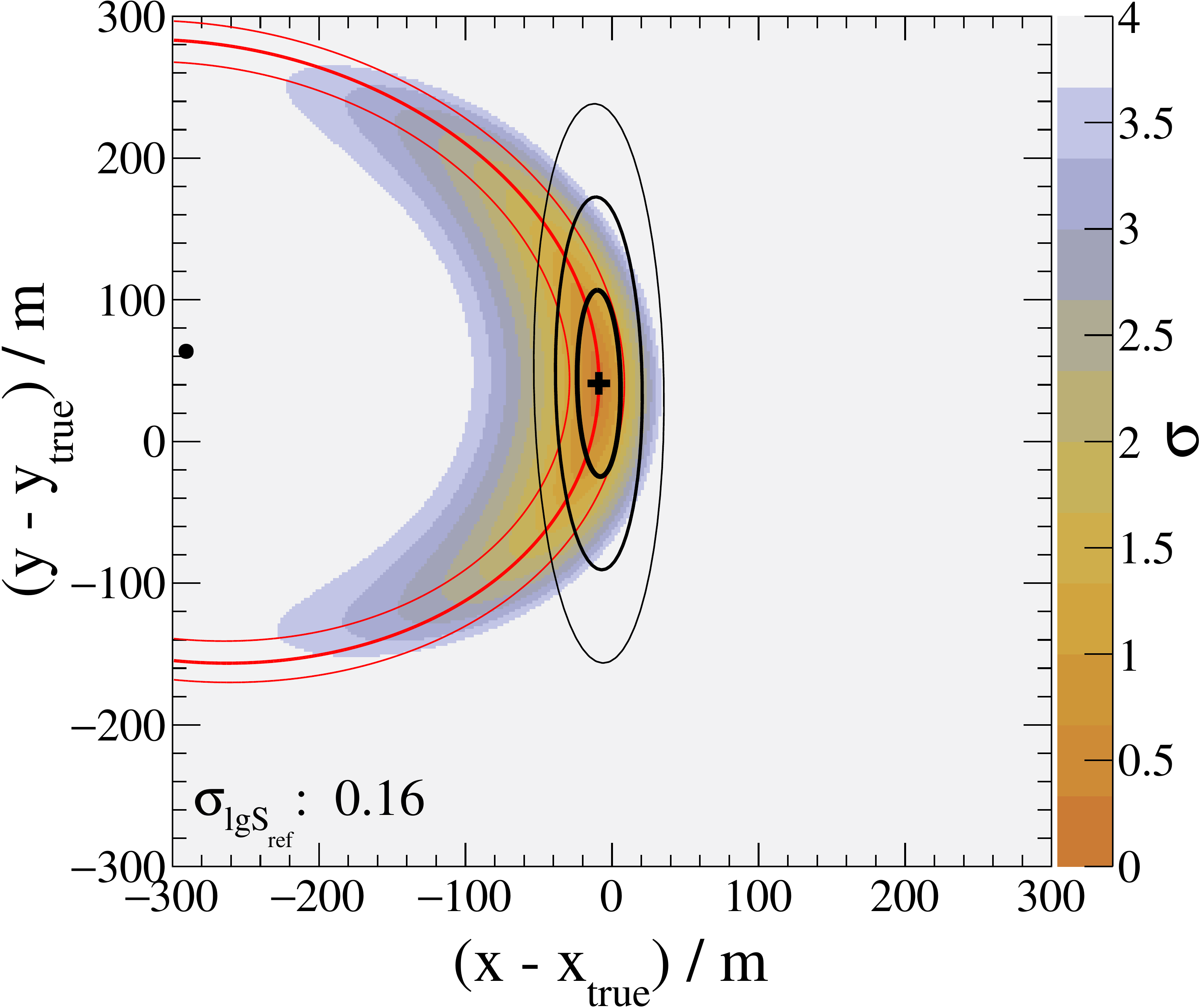}\\
\vspace{0.5cm}
\includegraphics[width=.49\textwidth]{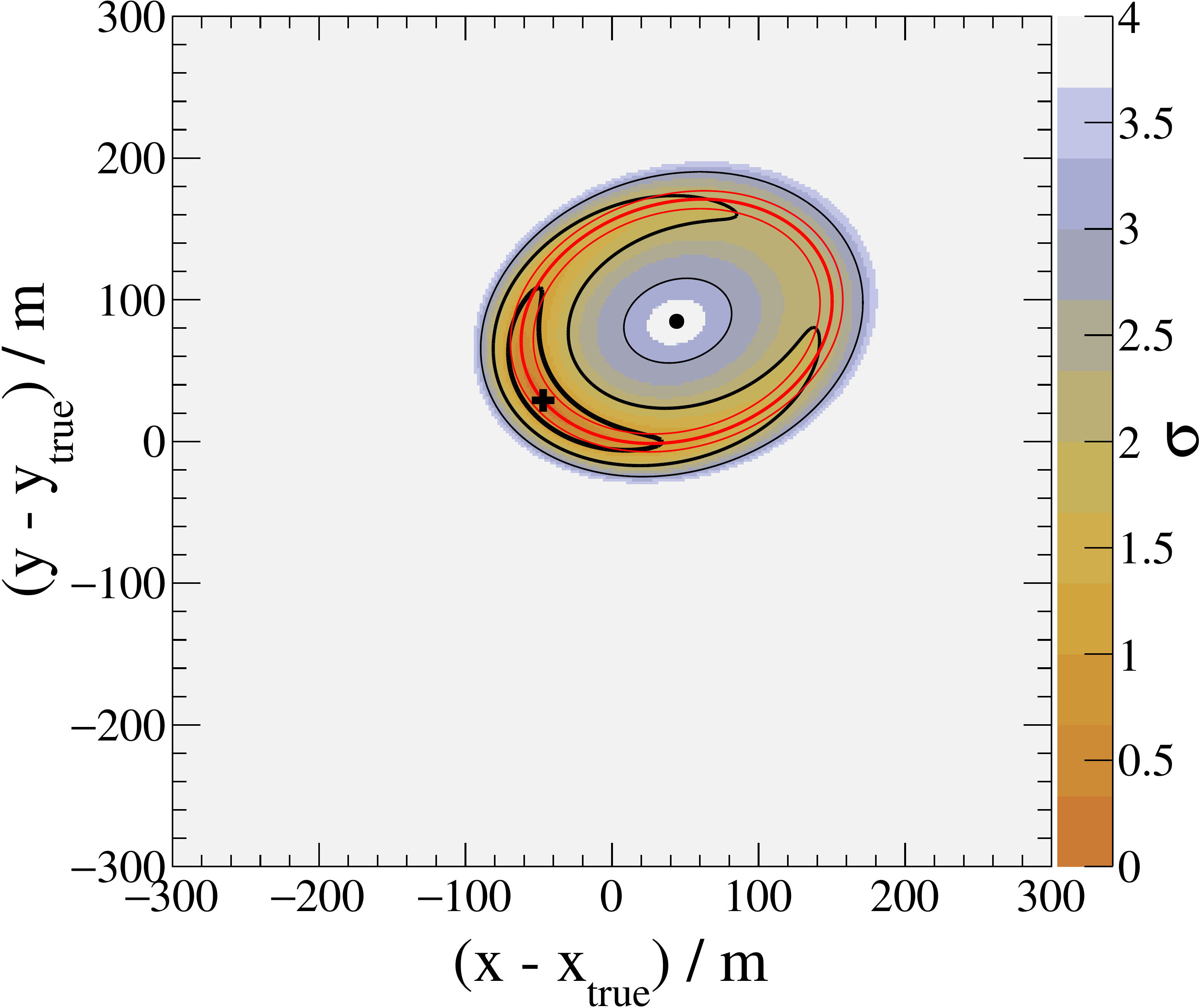}
\hfill
\includegraphics[width=.49\textwidth]{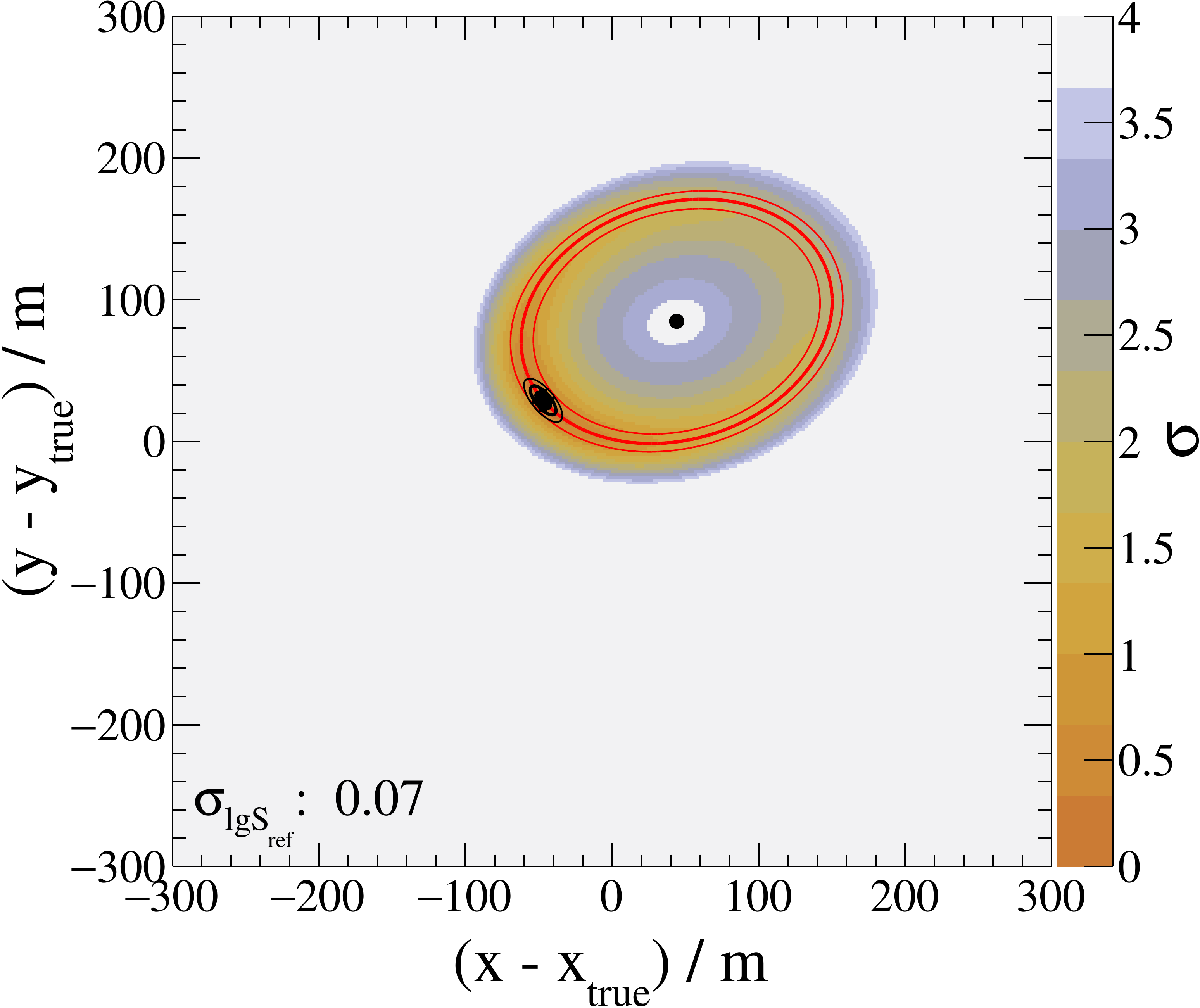}
\caption{\label{fig:llhspacecart} The LLH-space for an event with a zenith angle of $40^\circ$ is shown. The color scale indicates the confidence interval of a given impact point being correct. The location with the largest LLH value is indicated with a black cross and the location of the detector with the largest signal is indicated by the black circle. The red lines are the contours of reconstructed \lgsref values with respect to that of the largest LLH value, $\lgsref / \lg S_{\rm best} = \{0.8, 1, 1.2\}$. The black lines show the 1$\sigma$, 2$\sigma$, and 3$\sigma$ contours. In the left panels, these contours represent the true ones while those in the right panel correspond to the ellipses given by the covariance matrix from MINUIT. The top and bottom panels show events where the shower axis is 227\,m and 84\,m from the shown detector location, respectively. In the panels on the right, the estimated uncertainty of the shower size, $\sigma_{\lgsref}$, is shown in the lower left corner.}
\end{figure}

\subsection{LLH-space for air-shower events}

Using the LLH defined above, the difficulty for a numerical minimization routine to estimate the uncertainties for an air-shower reconstruction can be made evident.
Two example events of air showers with zenith angles of $40^\circ$ are shown in \cref{fig:llhspacecart}.
Each panel shows the results of a scan of the potential impact point locations in $(x, y)$ i.e., along the ground.
At each location, the impact point and arrival direction were held fixed and only the value of \lgsref was left free.
The LLH space is seen to wrap around the detector with the largest signal, indicated by the black dot.
Exemplified by the 1$\sigma$, 2$\sigma$, and 3$\sigma$ contours, the LLH-space is stretched along the red lines of constant reconstructed \lgsref, which also encircle the detector with the largest signal.
This curvature is expected, both for the LLH and the \lgsref contours since the \LDF is dependent solely on $r$.

For an impact point location sufficiently far from a detector (e.g., top panels of \cref{fig:llhspacecart}), the 1$\sigma$ LLH-contour in Cartesian coordinates can be approximately described by an ellipse.
In this case, the second-derivative matrix calculated by MINUIT is sufficient to estimate the uncertainty.
Both the 1$\sigma$ LLH-contour and estimated ellipse reside inside the $\lg\sref / \lg S_{\rm best} = \{0.8, 1.2\}$ lines and are in agreement with the estimated uncertainty from MINUIT of $\sigma_{\lgsref} \simeq 0.16$.

The bottom panels highlight the problem of estimating the uncertainty in \lgsref.
In this case, the impact point is close enough to a detector that the highly-wrapped LLH-space is not sufficiently parabolic in Cartesian coordinates for the second-derivative calculation to provide a stable estimate of the uncertainty in \lgsref.
The curvature about the detector is degenerate with a gradient of the LLH-space and the relative uncertainty of the shower size is estimated to be $\sigma_{\lgsref} \simeq 0.07$ (bottom-right panel) even though the true 1$\sigma$ contour (bottom-left panel) shows that the uncertainty is at least 20\%.
For an air-shower experiment, this creates a erroneous result in the estimated uncertainties on \lgsref depending on the impact point.

\begin{figure}[tbp]
\centering 
\includegraphics[height=7.1cm]{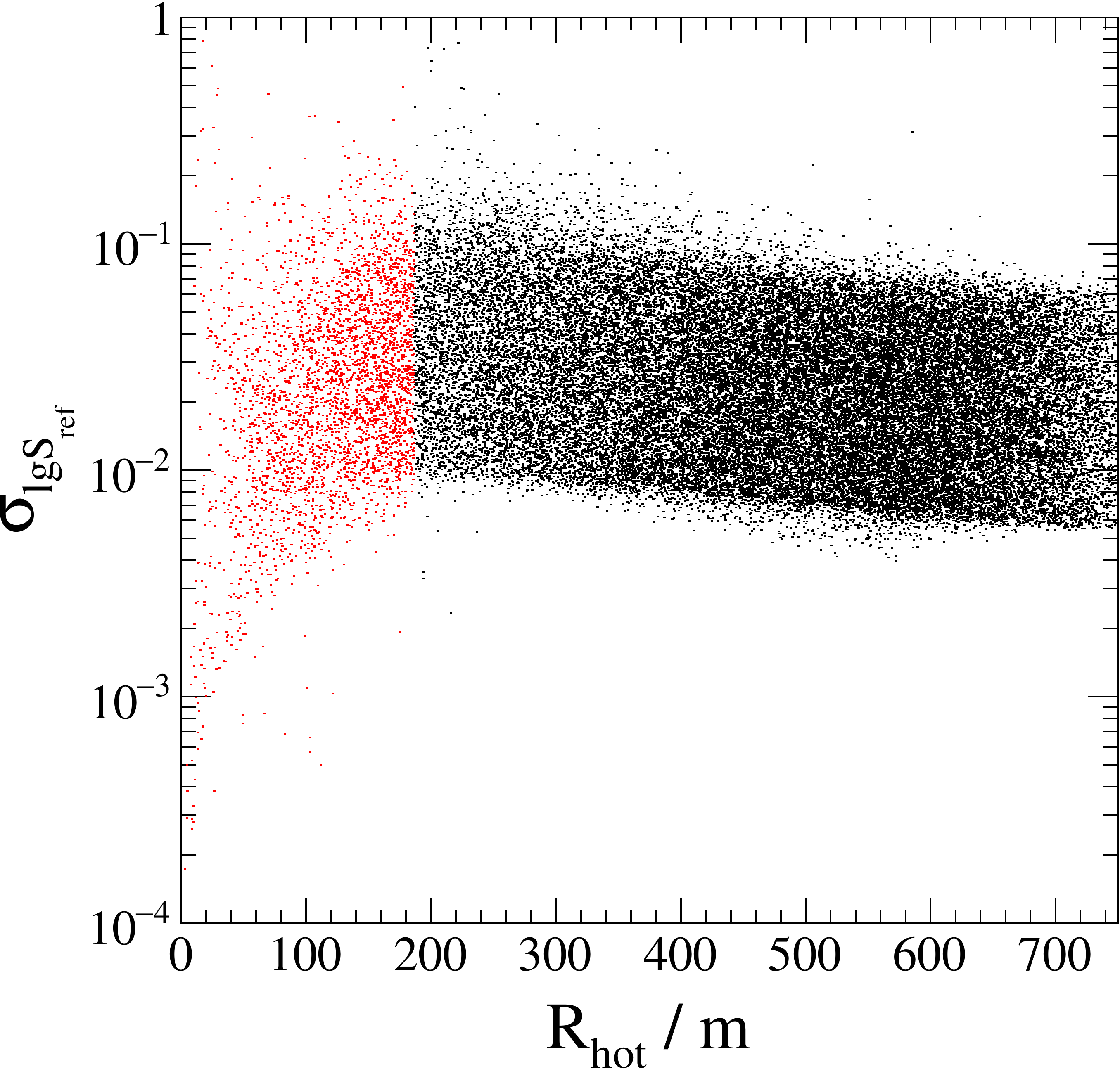}
\hfill
\includegraphics[height=7.1cm]{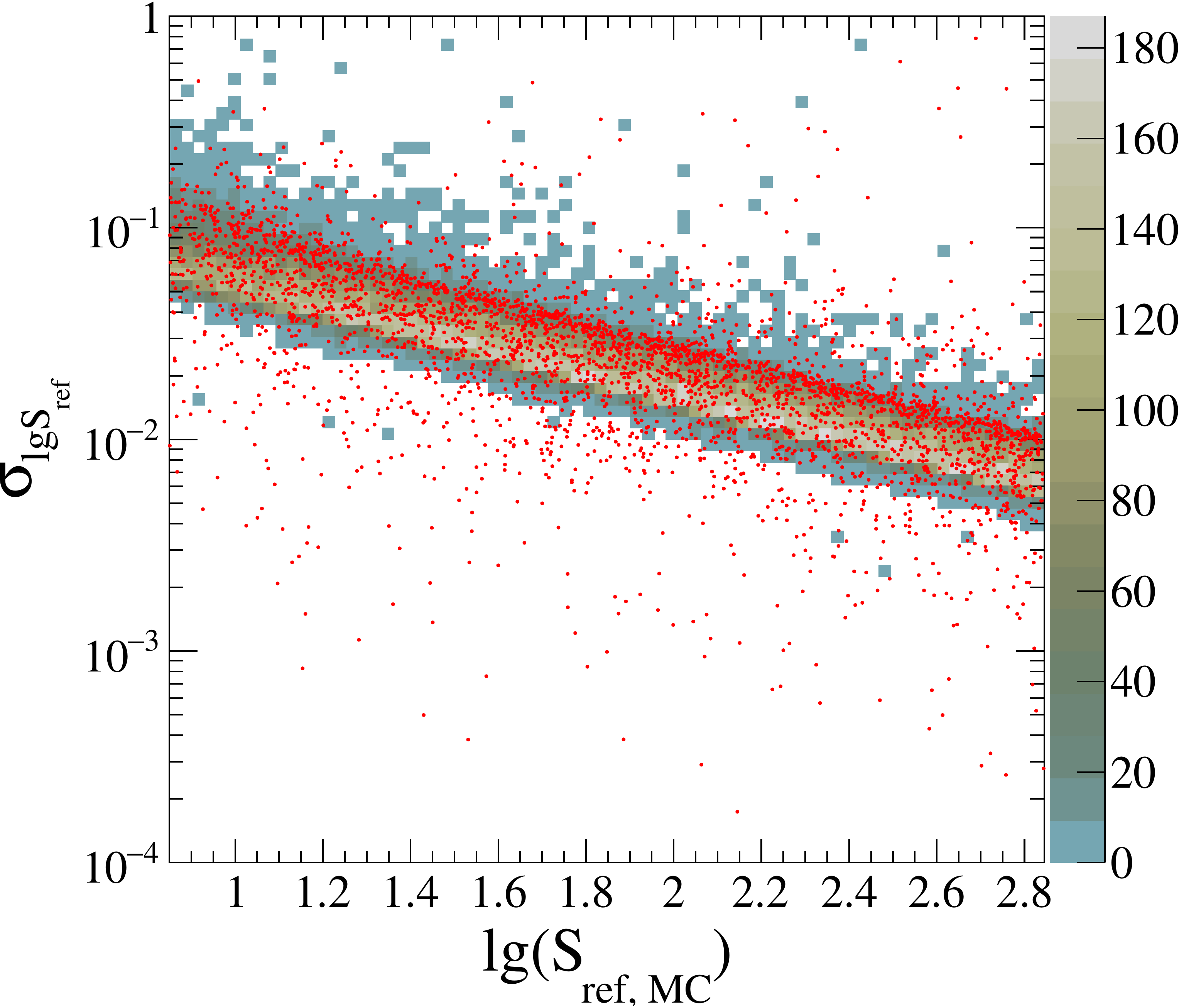}\\
\caption{\label{fig:sref_err_cart_hex} Left: Estimated relative uncertainty of the shower-size, as calculated using MINUIT, for 50\,000 simulated air showers. This distribution is shown as a function of the distance of the detector with the largest signal from the shower axis. The values in red are those with $\rhot < 200$\,m. Right: Histogram of the uncertainty for simulations with $\rhot \geq 200$\,m as a function of the true shower size \lgsref. The red points are the same events that are shown in red in the left panel.}
\end{figure}
To study the phase space of impact points where this effect is most prominent, 50\,000 events were simulated and reconstructed. Random uniform azimuth angles were chosen and the impact points were uniformly picked within the boundaries highlighted in~\cref{fig:array}. The impact point $(x,y)$  and \lgsref were left as free parameters during the minimization.
The errors on \lgsref, as estimated by the MINUIT algorithm, are shown in \cref{fig:sref_err_cart_hex}.
As seen in the left panel, the relative uncertainty decreases with increasing the distance from the shower axis, \rhot, to the detector with the largest signal.
However, for small values of \rhot, shown in red, there is an opposite trend and the uncertainties are up to an order of magnitude too small.
In the right panel, the distribution of events is shown versus shower size, again with small values of \rhot shown in red.
The underestimated shower uncertainties are seen to occur at all shower-size values meaning that this effect is independent of the number of triggered detectors.

\section{Fitting in log-polar coordinates}
\label{sec:polar_coords}

\subsection{Log-polar coordinates}
\begin{figure}[tbp]
\centering 
\includegraphics[width=.6\textwidth]{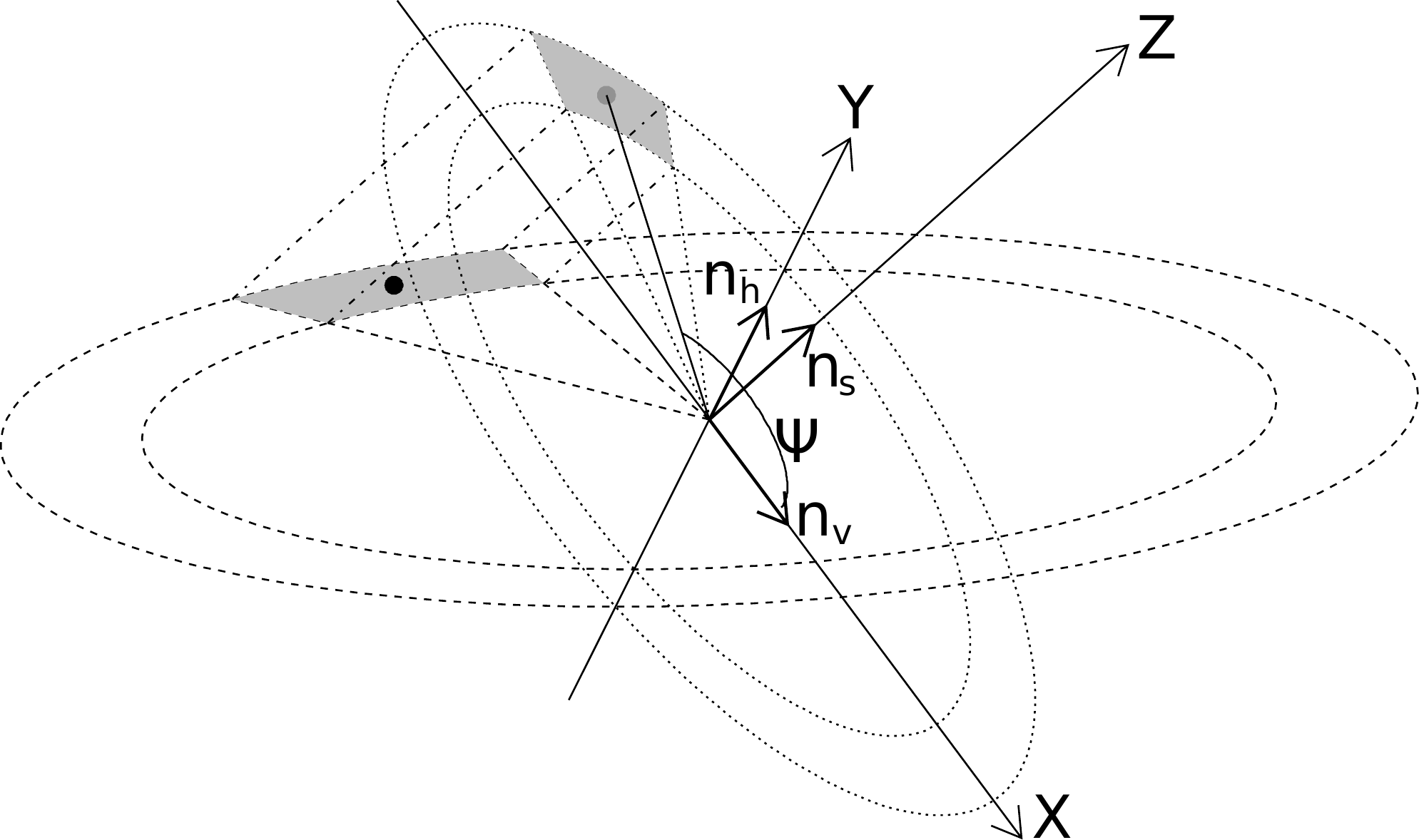}
\caption{\label{fig:showerfront}Diagram of the shower frame coordinate system: the $Z$ axis is along the shower axis $\mathbf{n}_{\rm s}$, the $X$ axis is along $\mathbf{n}_{\rm v}$ in the vertical plane going through the shower axis, and the $Y$ axis is along $\mathbf{n}_{\rm h}$ in the horizontal plane. In this coordinate system, $\psi$ is used to describe the polar angle about the shower axis, see~\cref{eq:coordsys}.}
\end{figure}
\begin{figure}[tbp]
\centering 
\includegraphics[width=.49\textwidth]{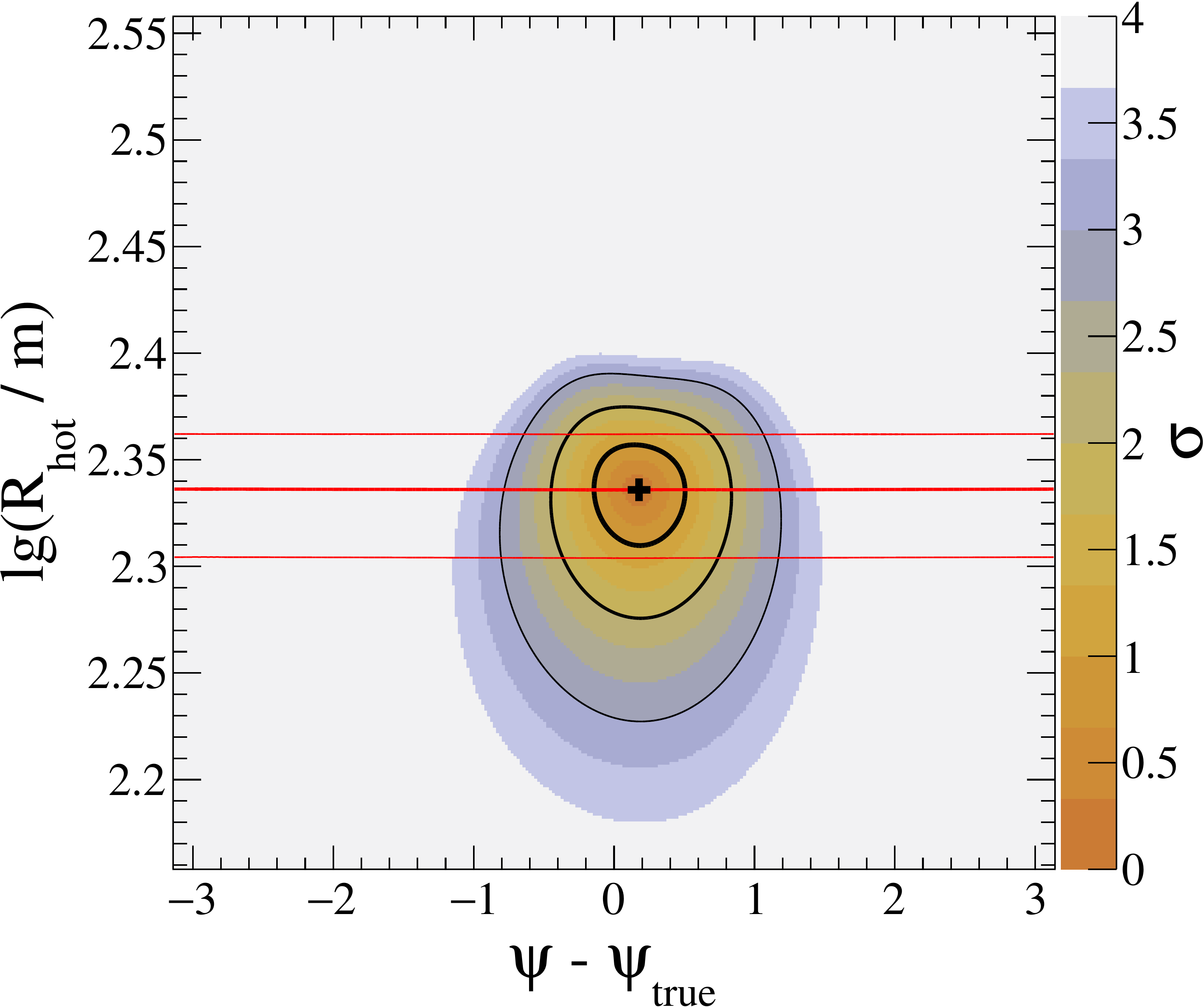}
\hfill
\includegraphics[width=.49\textwidth]{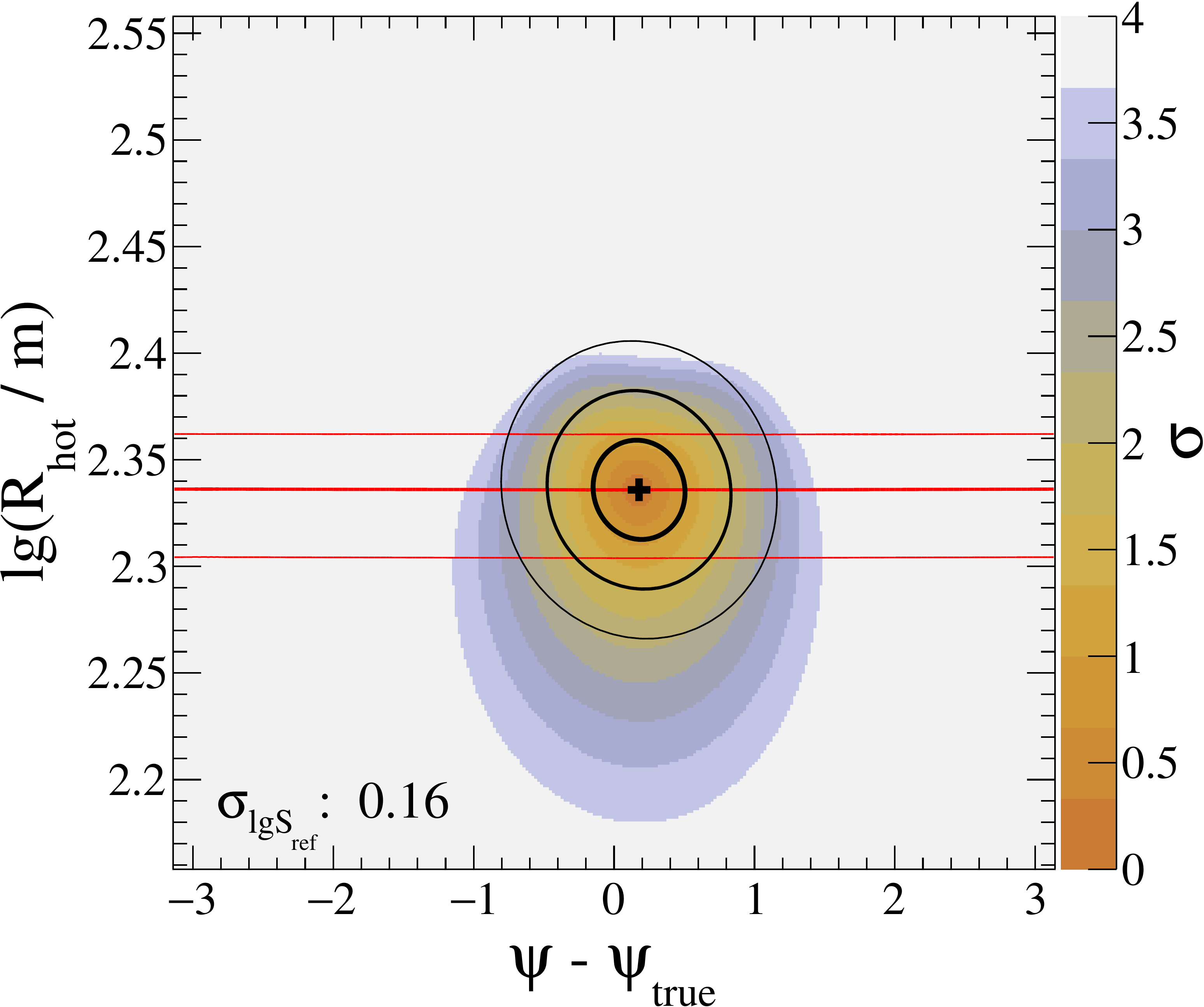}\\
\vspace{0.5cm}
\includegraphics[width=.49\textwidth]{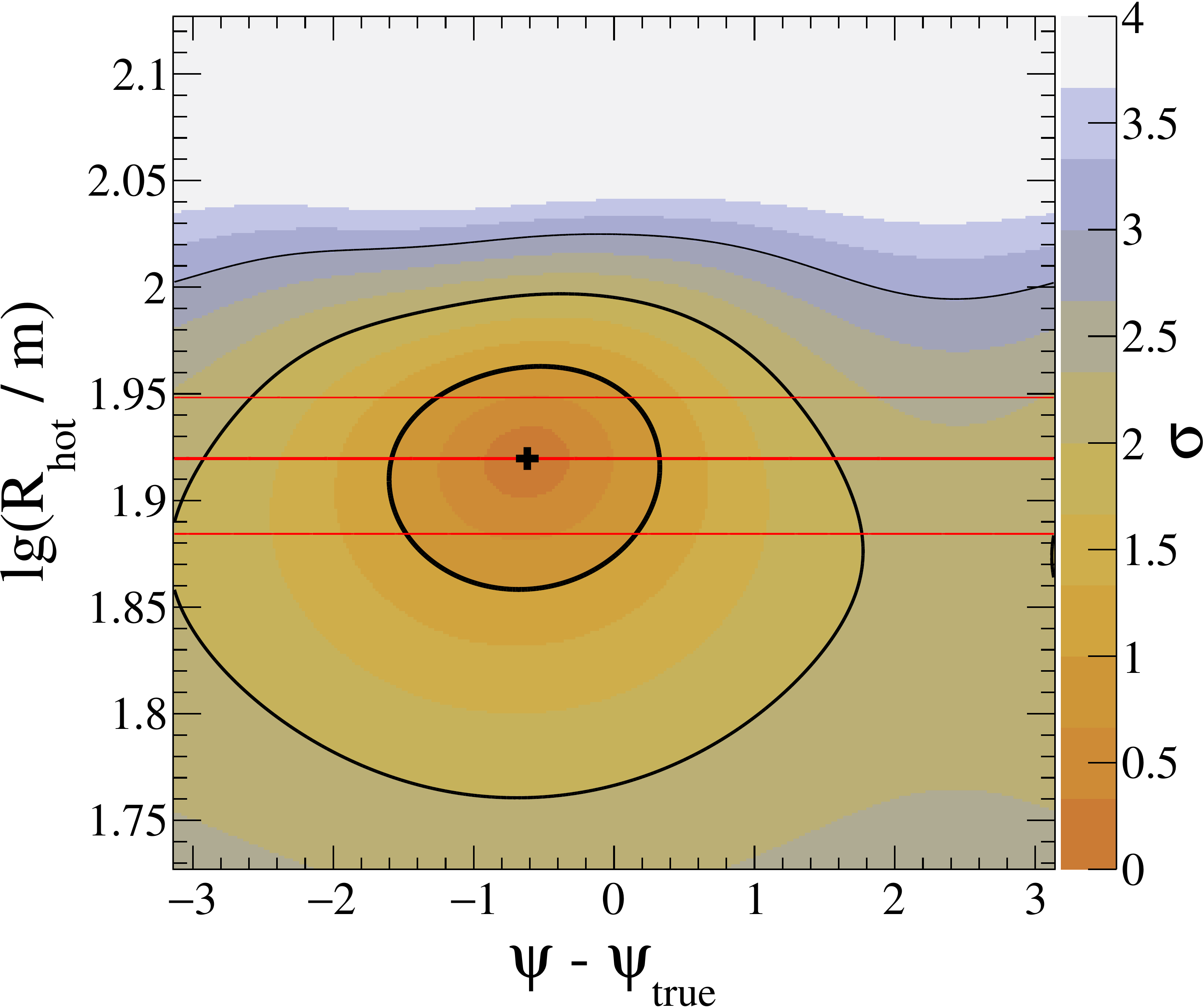}
\hfill
\includegraphics[width=.49\textwidth]{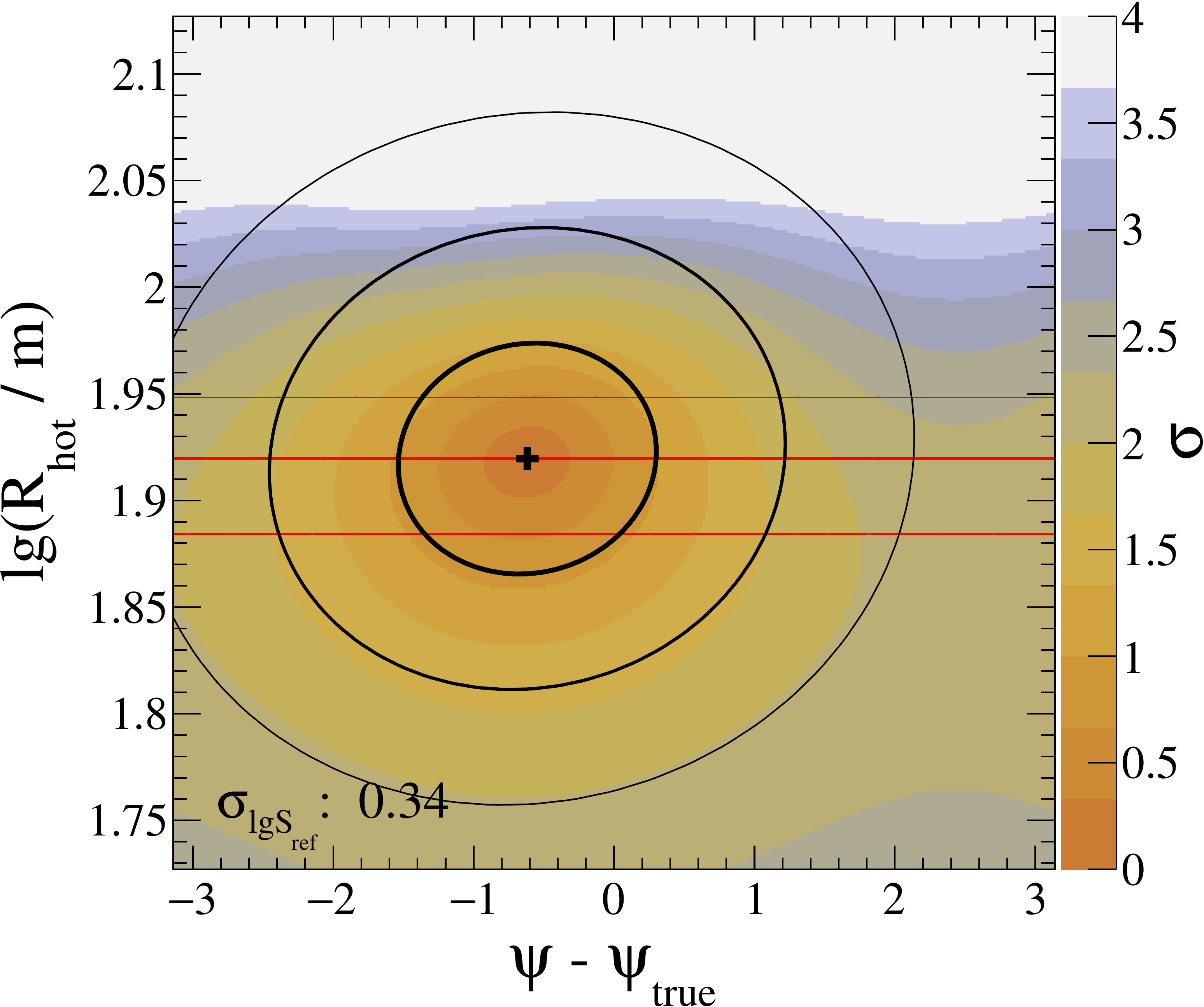}
\caption{\label{fig:llhspacepolar} The top and bottom panels show the LLH-space for the same two events as those in \cref{fig:llhspacecart}, this time calculated in the log-polar coordinate system. The true 1$\sigma$, 2$\sigma$, and 3$\sigma$ LLH contours are given in the left panels while the right panels are the respective ellipses from the covariance matrix from MINUIT.}
\end{figure}
The issue described above is ultimately related to the non-parabolic nature of the LLH-space in terms of the parameters that are given to the minimizer, namely the Cartesian location of the impact point.
Instead we propose a more natural coordinate system for the task of reconstructing the impact point and size.
To develop the new parameters, we first define the Cartesian \emph{ground coordinate system} wherein an air shower has zenith and azimuth angles of $\theta$ and $\phi$, respectively, and the impact point is described by the $(x,y)$ intersection of the shower axis with the ground.
This system is depicted in~\cref{fig:showerfront} wherein a sampling area, such as that around the black point, is the projection onto the ground, along the $Z$ axis, of a region in the shower plane $(X,Y)$.
The shower frame is thus defined by $\mathbf{n}_{\rm s}$ and two perpendicular axes, for example an horizontal one, $\mathbf{n}_{\rm h}$, and the other one, $\mathbf{n}_{\rm v}$, into the vertical plane containing $\mathbf{n}_{\rm s}$. 
We consider in the following the shower-front coordinates obtained from the ground ones as,
\begin{subequations}\label{eq:coordsys0}
\begin{align}
    X_{\rm sc} &= (\Delta x_{\rm hot} \cos \phi + \Delta y_{\rm hot} \sin\phi) \cos \theta \label{eq:x_sc},\\
    Y_{\rm sc} &= -\Delta x_{\rm hot} \sin \phi + \Delta y_{\rm hot} \cos \phi \label{eq:y_sc}.
\end{align}
\end{subequations}
Here, $\Delta x_{\rm hot}$ and $\Delta y_{\rm hot}$ define a Cartesian location in the ground coordinate system with respect to the detector with the largest signal.
The values of $X_{\rm sc}$ and $Y_{\rm sc}$ are the 2D Cartesian locations in the shower coordinate system.  
To reconstruct the air-shower size, we then propose the ``log-polar'' coordinate system defined by the transformation into the shower plane,
\begin{subequations}\label{eq:coordsys}
\begin{align}
    R &= \sqrt{X_{\rm sc}^2 + Y_{\rm sc}^2} \label{eq:lg_r},\\
    \Psi &= \arctan2(Y_{\rm sc}, X_{\rm sc}) \label{eq:phi}.
\end{align}
\end{subequations}
The new variables for reconstructing air showers are given by $\lg R$ and $\Psi$, the log-radius and polar angle about a line that is parallel to the shower axis and passes through the detector with largest signal.
In addition, \lgsref, the (decimal) logarithm of \sref, is considered as the third free parameter.
The choice of anchoring this coordinate system to the detector with the largest signal is motivated by the curvature of the LLH-space, as previously seen in \cref{fig:llhspacecart}.

\subsection{Results}
\begin{figure}[tbp]
\centering 
\includegraphics[height=7.1cm]{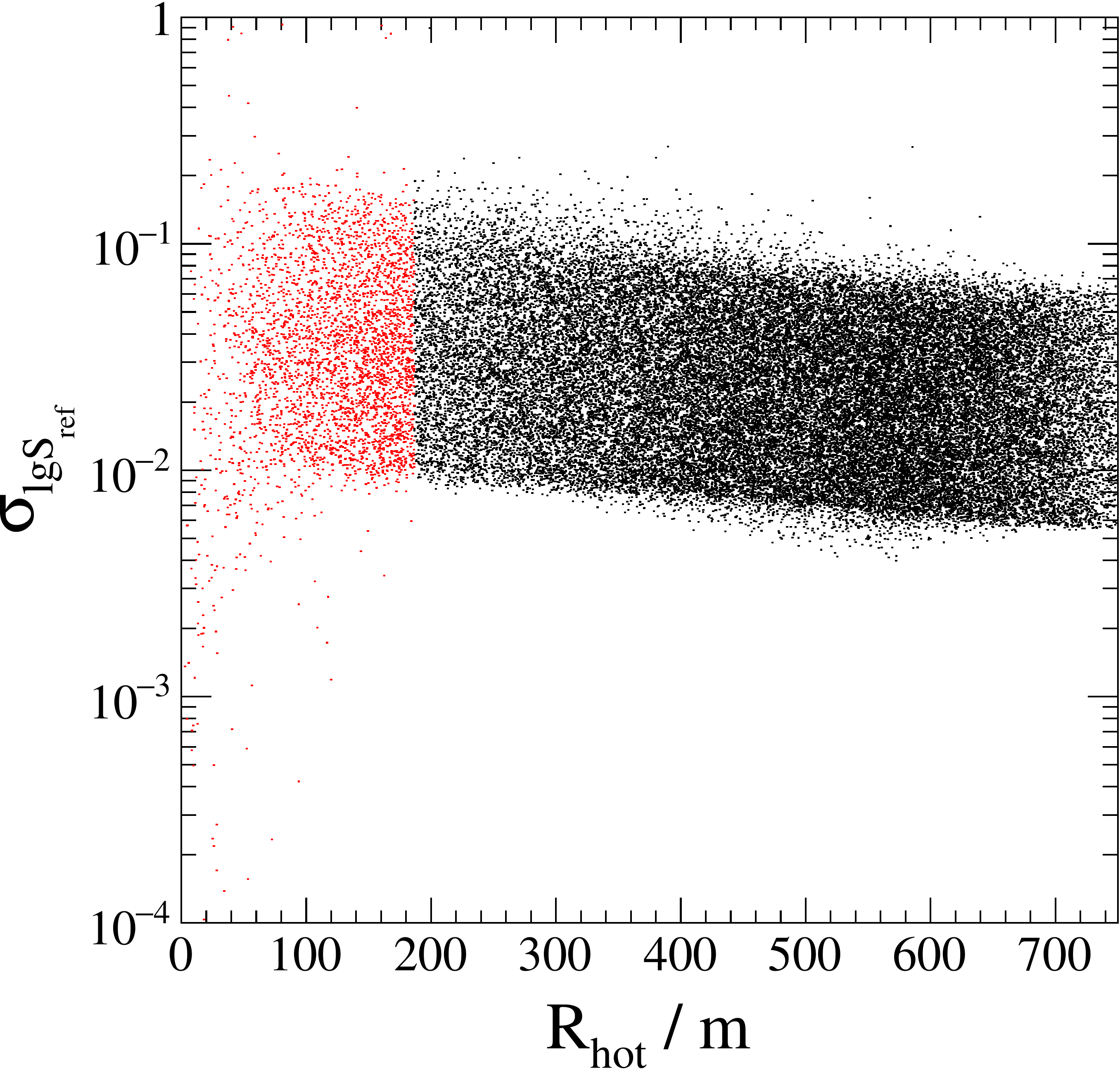}
\hfill
\includegraphics[height=7.1cm]{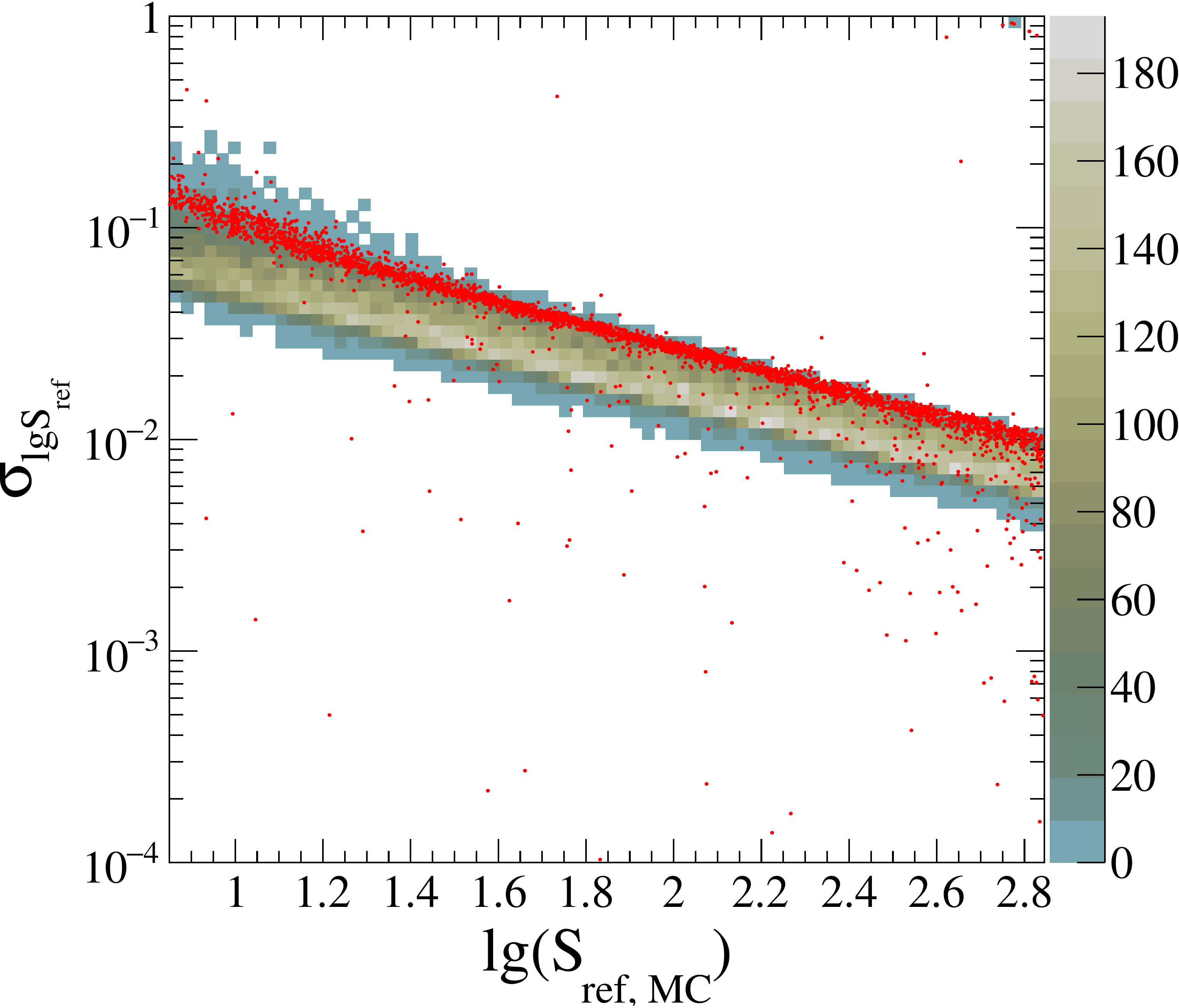}\\
\caption{\label{fig:sref_err_polar_hex} The same distributions as shown in \cref{fig:sref_err_cart_hex}, using the same 50\,000 Monte Carlo trials, but calculating the shower-size uncertainty using the log-polar coordinates defined in \cref{eq:coordsys}.}
\end{figure}
In \cref{fig:llhspacepolar}, the same two example events are shown in the log-polar coordinate system defined above.
For both sample events, several nice properties of the LLH space are seen.
Firstly, in this space, the lines of constant reconstructed \lgsref are almost independent of the $\Psi$ coorduncinate.
This is beneficial as this means that the correlation between $\Psi$ and \lgsref is small, regardless of the direction of the shower propagation, which is helpful for attaining a more robust estimate of the shower-size uncertainty.
Secondly, the $1\sigma$ contours are more elliptical meaning that the assumption when using the Hessian matrix method is more valid and will typically return a better error estimate.

We repeated the study shown in \cref{fig:sref_err_cart_hex} on the same 50\,000 simulated events but using the log-polar coordinate system described above.
The results are shown in \cref{fig:sref_err_polar_hex}.
In this case, the uncertainty estimation is much more well behaved with most of the events with $\rhot < 200~$m appearing now in a narrow band as a function of \sref.
This band occurs because shower axes that pass very near to a detector are necessarily maximally far from all other detectors.
On one hand, this will produce less triggered detectors per event.
On the other hand, for the detectors with signal, the uncertainties, which scale like $\sqrt{S(r)} / S(r) = 1/\sqrt{\LDF(r)} \propto r^{-\beta / 2}$, will be larger than average for a given shower size.
Although the ``hottest'' detector has the smallest relative uncertainty, its weight in the fit is mitigated compared to that of other detectors.
This is because a small shift $\delta r$ in the impact point results in a relative change $\beta\,\delta r/r$ for $S$ and is proportional to $r^{-\beta/2-1}$ in units of $\sigma(S)$. 
The fitting procedure can thus change the expected signal of the hottest detector at a low cost through a change of the impact point, without affecting the expectation for the other detectors.
All in all, these effects produce a less constrained LLH space and results in a larger uncertainty in the shower size.
While there are a few outliers, these make up only $\sim$\,0.1\% of all events.

\subsection{Dependence on array geometry}
\begin{figure}[tbp]
\centering 
\includegraphics[height=7.1cm]{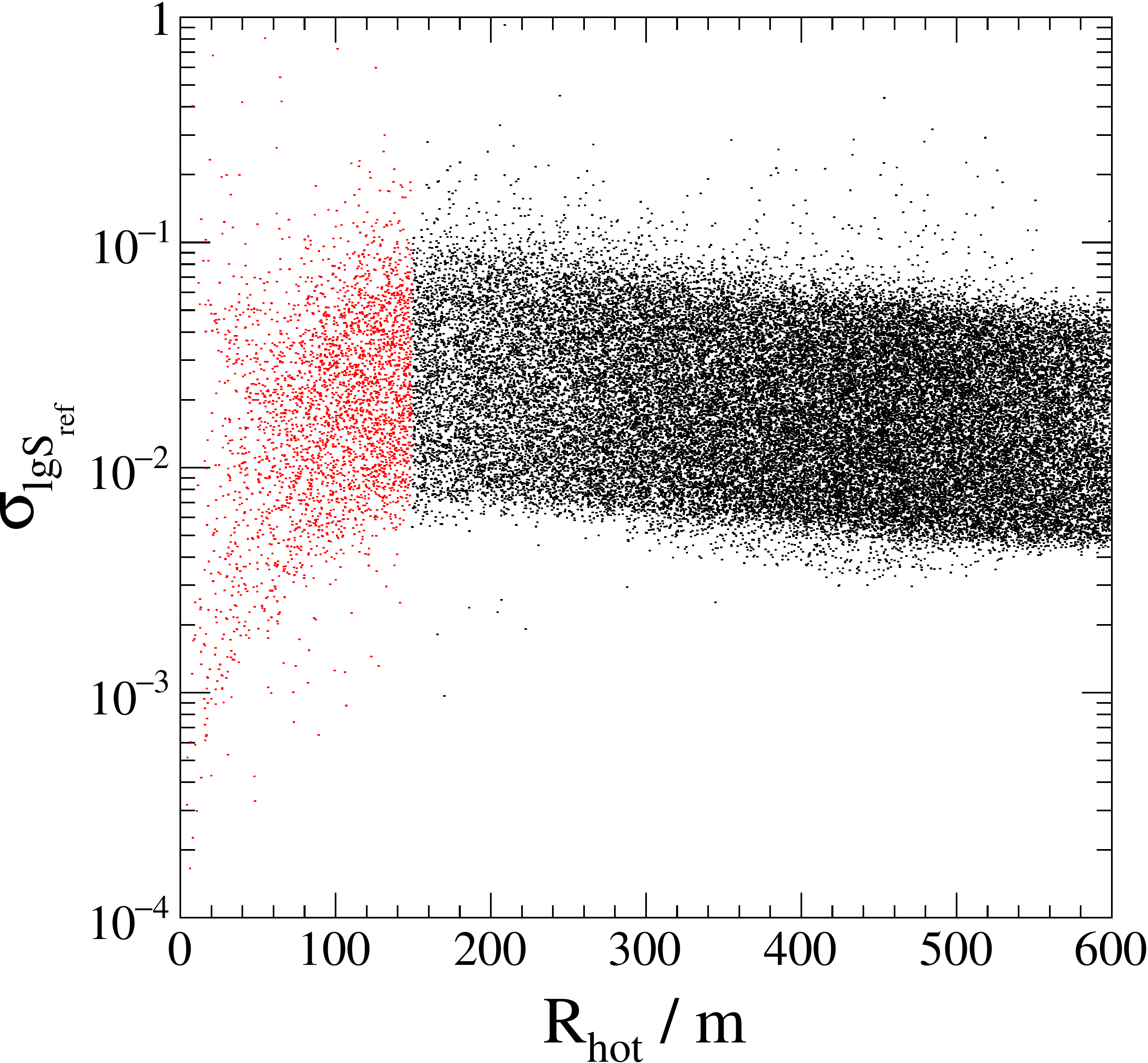}
\hfill
\includegraphics[height=7.1cm]{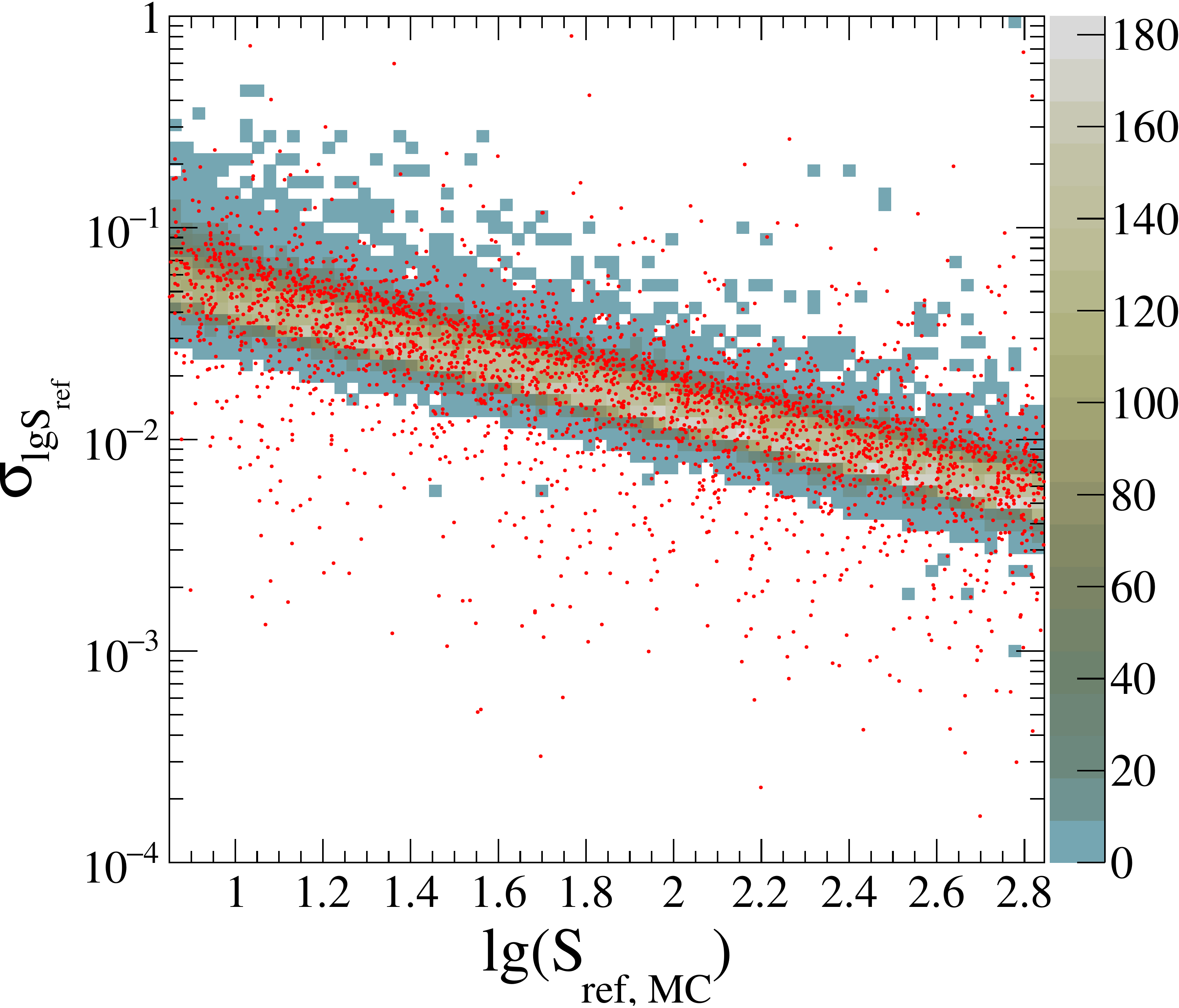}\\
\vspace{0.5cm}
\includegraphics[height=7.1cm]{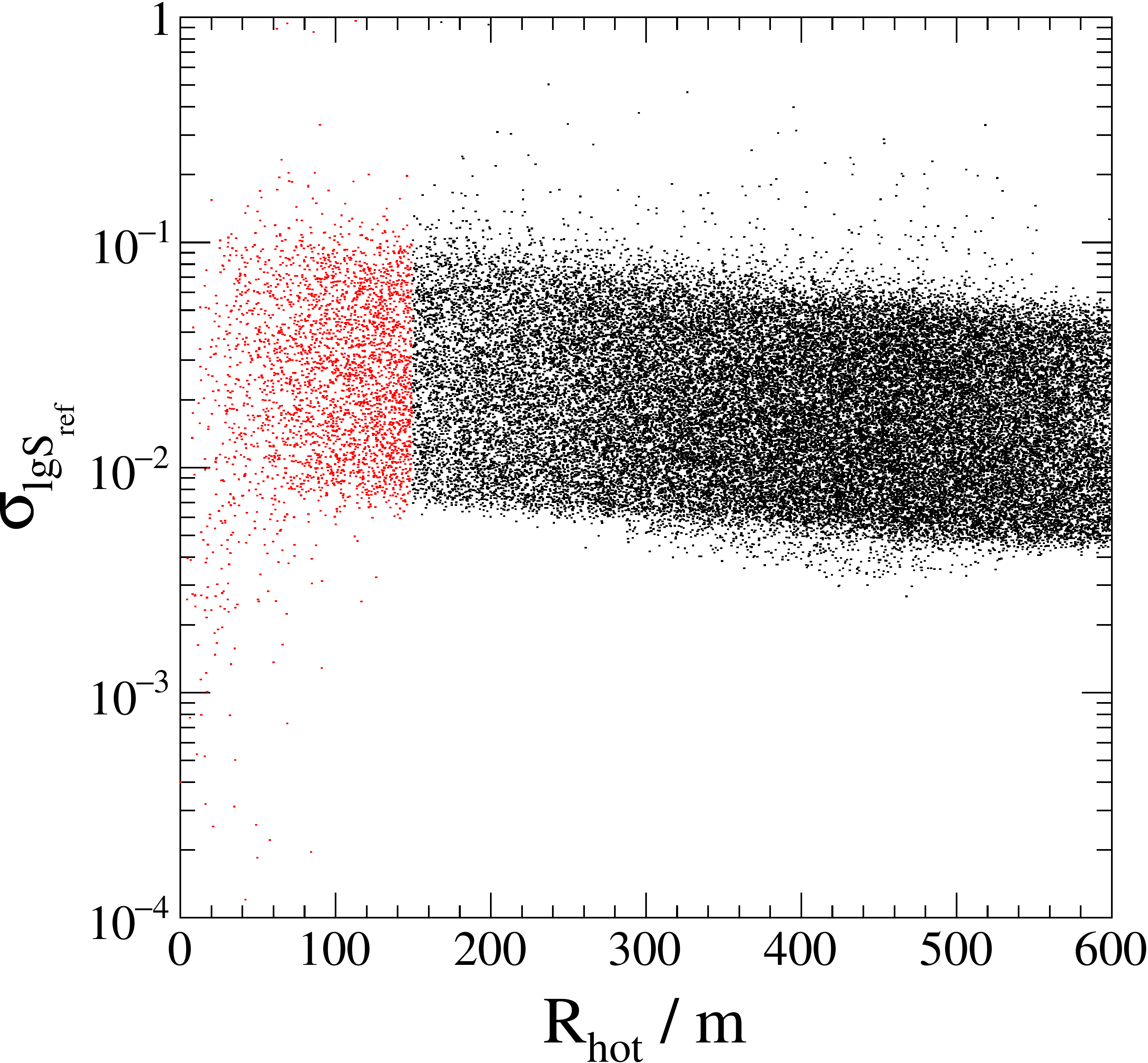}
\hfill
\includegraphics[height=7.1cm]{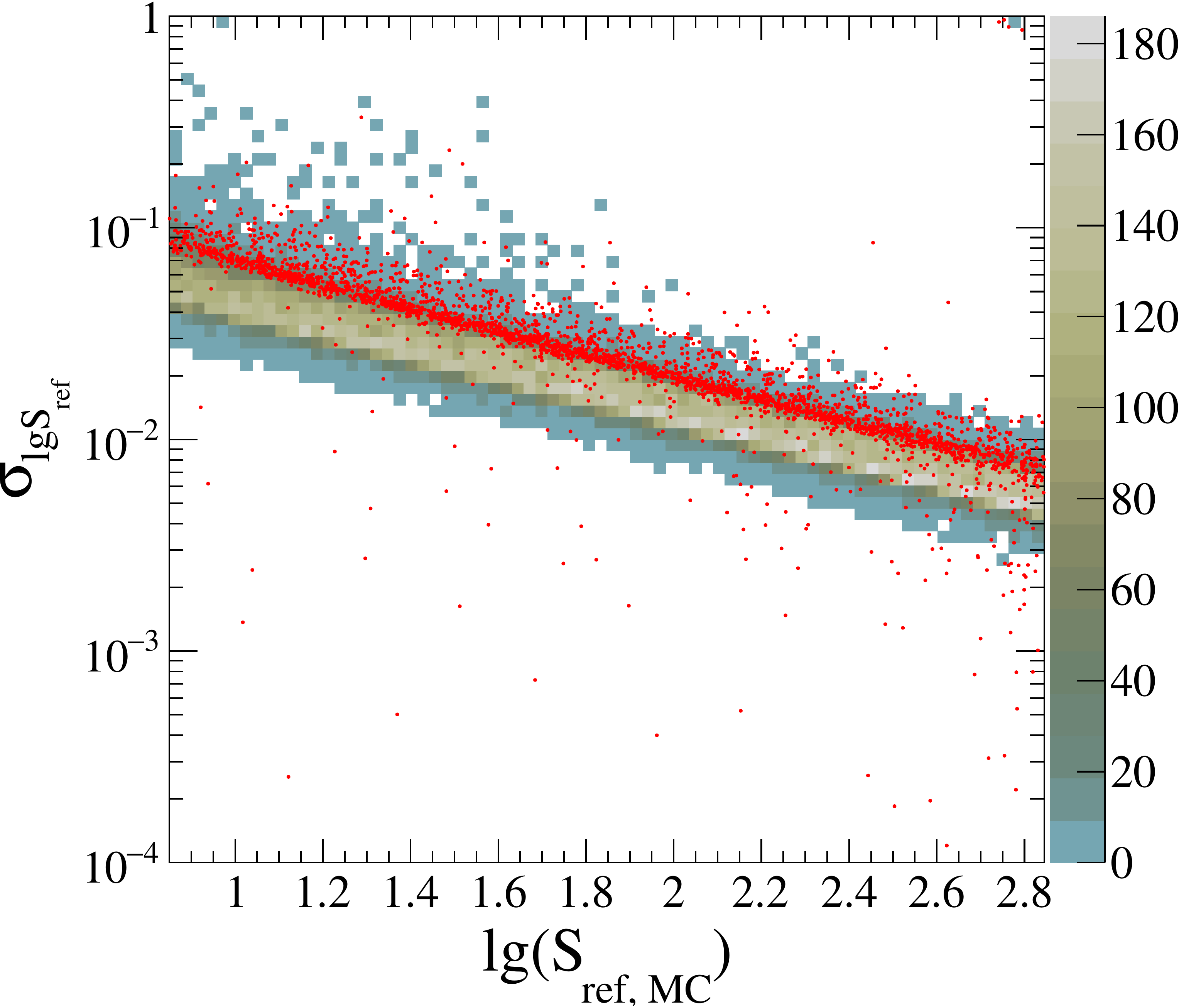}
\caption{\label{fig:ta_spacing} The distributions of the relative estimated shower-size uncertainty are shown for the Telescope Array detector layout. In this case, the red points are defined by events with $\rhot < 100$\,m. The top panels correspond to reconstructions using Cartesian coordinates (as in \cref{fig:sref_err_cart_hex}) while the bottom panels correspond to reconstructions using the log-polar coordinates (as in \cref{fig:sref_err_polar_hex}).}
\end{figure}
Given that this work is focused on a geometric issue, it is important to understand how the layout of the array ultimately impacts the shower-size estimation.
We contrast the main results shown above after changing to the surface detector layout of Telescope Array, a square grid with 1200\,m spacing between detectors, as shown in the right panel of \cref{fig:array}.
The results are shown in \cref{fig:ta_spacing}.

For the more dense detector arrangement, more detectors are triggered for a given shower size and the precision on \lgsref is better.
However, it is clear that the same issue is present for this array configuration, as seen in the top panels.
The radius at which the shower size uncertainty undergoes an inflection is even in a similar location given the length scales involved, occurring at $\rhot / d_{\rm spacing} \simeq 0.125$ where $d_{\rm spacing}$ is either 1500\,m or 1200\,m for the respective array layouts.
Ultimately, a significant improvement is still observed by using the log-polar coordinates in the air shower reconstruction, as seen in the bottom panels.

\subsection{Effects of saturation}
In measured air-shower data, detectors near the shower axis can saturate, providing an unreliable estimate of the signal content at that point.
Since it is exactly these types of showers that result in non-quadratic LLH contours, it is worthwhile to consider the way that saturated stations may impact the results presented above.
While the non-linear effects that occur near the upper end of the dynamic range of a detector will be hardware-dependent, we show a few limiting cases for handling saturated detectors during reconstruction.

In the first case, the underlying signal that would have been measured in the absence of saturation is recovered post-hoc.
Whether by analytical methods or using machine learning, various techniques have been employed to estimate the would-be signal in saturated detectors, e.g.~\cite{Yue:2003fh, Veberic:2013suc, Liu:2019dvt}.
In the best case, the signal that would have been measured is precisely estimated with the result that effectively nothing has changed with respect to the results shown above.
In the less ideal case, the recovery introduces an additional uncertainty, i.e. $\sigma(S) \longrightarrow \sigma(S) + \Delta(S)$, which can simply be taken into account in the reconstruction (\cref{eq:llh}).
This would widen the LLH contours, depending on the relative size of $\Delta(S)$, but would not remove the non-quadratic behavior close to the saturated detector.

\begin{figure}[tbp]
\centering
\includegraphics[width=.49\textwidth]{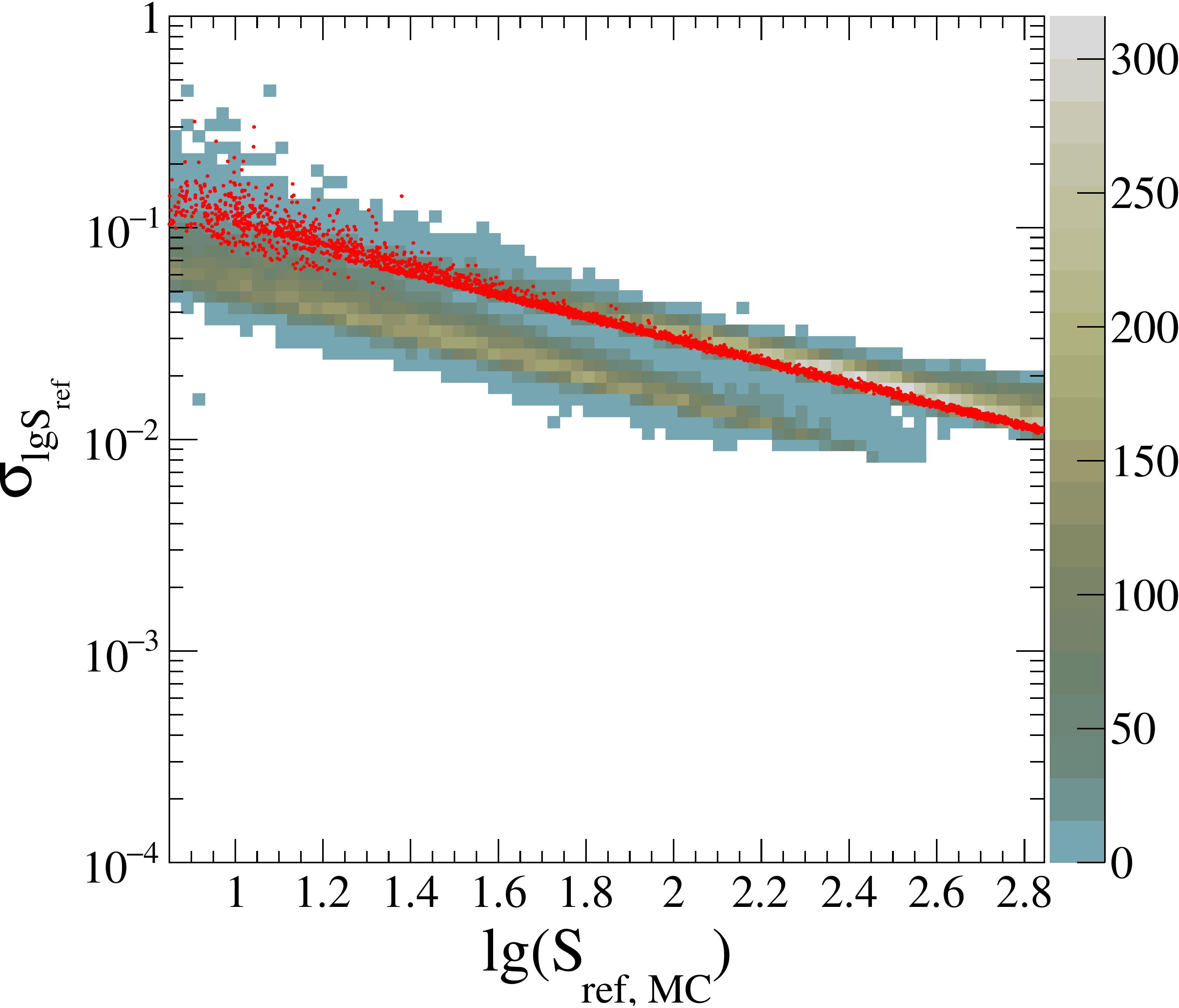}
\hfill
\includegraphics[width=.49\textwidth]{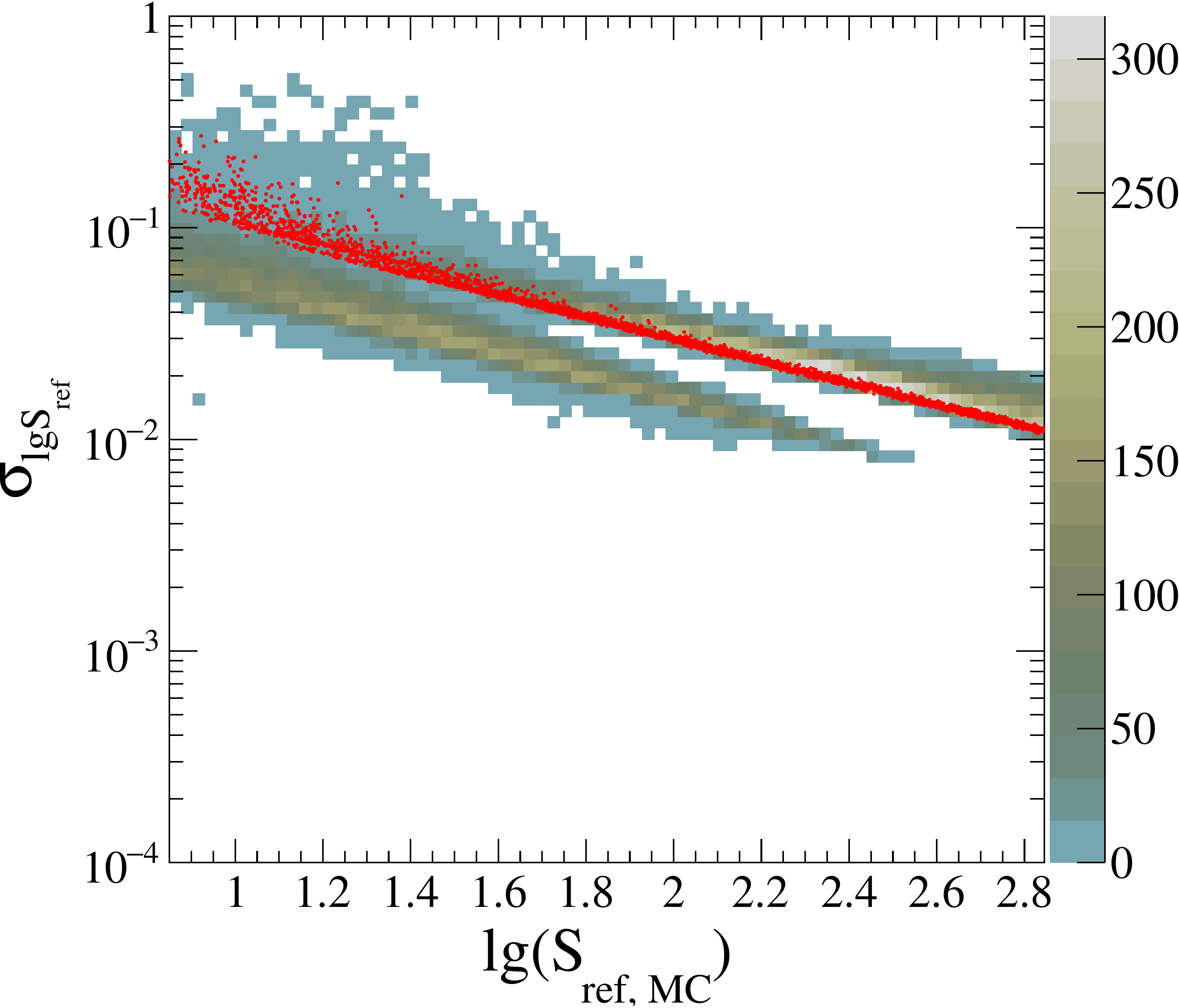}
\caption{\label{fig:saturation} Left: The uncertainty in the shower size, when using an additional term in the LLH to account for saturated stations (\cref{eq:LLH_sat_term}) when performing a reconstruction using Cartesian coordinates. Right: The same figure when ignoring saturated stations in during the reconstruction of the air-shower size. Both panels in this figure are directly comparable to~\cref{fig:sref_err_cart_hex}.}
\end{figure}

Alternatively, a corresponding term can be included in the likelihood function to describe the probability that a detector observes saturation given an expected signal.
In the case of ideal saturation, where signals only up to \ssat can be observed, the corresponding term to include in~\cref{eq:llh} is,
\begin{equation}
    \label{eq:LLH_sat_term}
    - 2 \sum_k^{\rm saturated} \ln\left(1 - {\rm erf}\left(\frac{\ssat - S_{k}}{\sqrt{2}\,\,\sigma(S_k)} \right) \right).
\end{equation}
This was tested within the framework of this toy model using $\ssat = 1000$ and the results are shown in the left panel of~\cref{fig:saturation}.
In this case, the Cartesian coordinates already present a good estimation of the shower size uncertainty.
This is a result of the rather non-restrictive nature of the additional term in~\cref{eq:LLH_sat_term} for which expected signals of 1500 and 15\,000 (i.e. larger and smaller \rhot values) give a probabilities that are indistinguishable from each other within typical numerical precision.
This term does not have the ``repulsive" effect that is observed in~\cref{fig:llhspacecart} and thus the LLH-space is already fairly quadratic.
However, this comes at the cost of roughly half an order of magnitude in statistical precision which can be important for performing a physics analysis on the highest-energy cosmic rays.

In the final case, the saturated detector is completely ignored during the reconstruction of the air-shower size and effectively treated as a hole in the array.
This was studied using a saturation threshold of 1000 and is shown in the right panel of~\cref{fig:saturation}.
Similar to the previous case, the lack of repulsion from the hottest detector does not warp the LLH space in Cartesian coordinates.
The estimations are again robust but at the cost of numerical precision on \sref.

Note that the exact details of the gap in precision between the events with and without saturation and where these distributions begin as a function of air-shower size depend on the \LDF shape, the true air-shower size, and the details of detector system.
Shower-to-shower fluctuations, ignored in this study, should also have impact.
However, it can be concluded that in the second and third cases, when the saturated detectors do not have any effective influence on the preferred core location, the use of Cartesian coordinates does not exhibit the biases in the uncertainty of \sref.
Finally, note that using the log-polar coordinate system produces equivalent uncertainty values for the second and third cases and is thus still a more robust system that can account for any of the methods described above.

\section{Application example}
\label{sec:application}

\begin{figure}[tbp]
\centering 
\includegraphics[width=.49\textwidth]{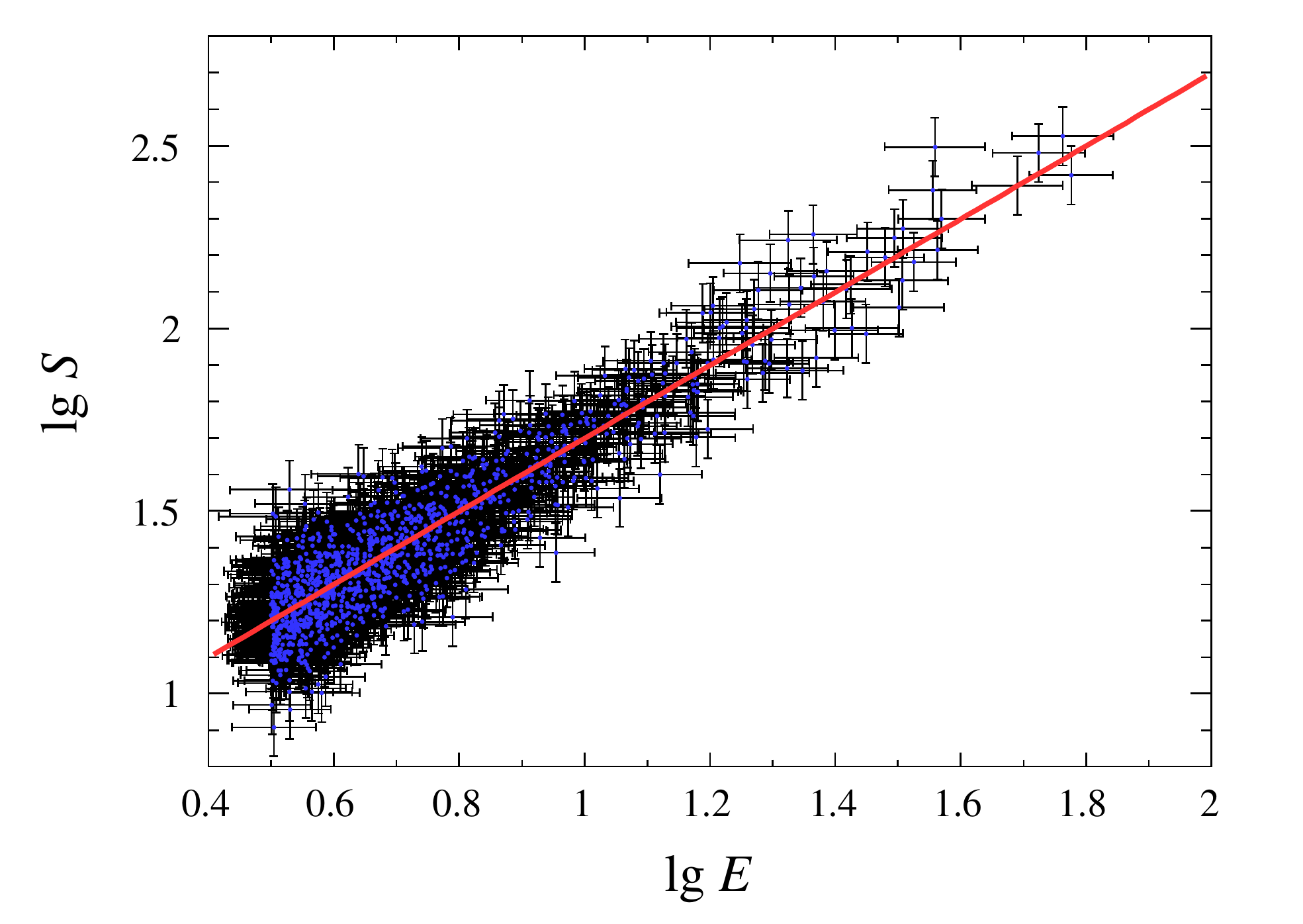}
\hfill
\includegraphics[width=.49\textwidth]{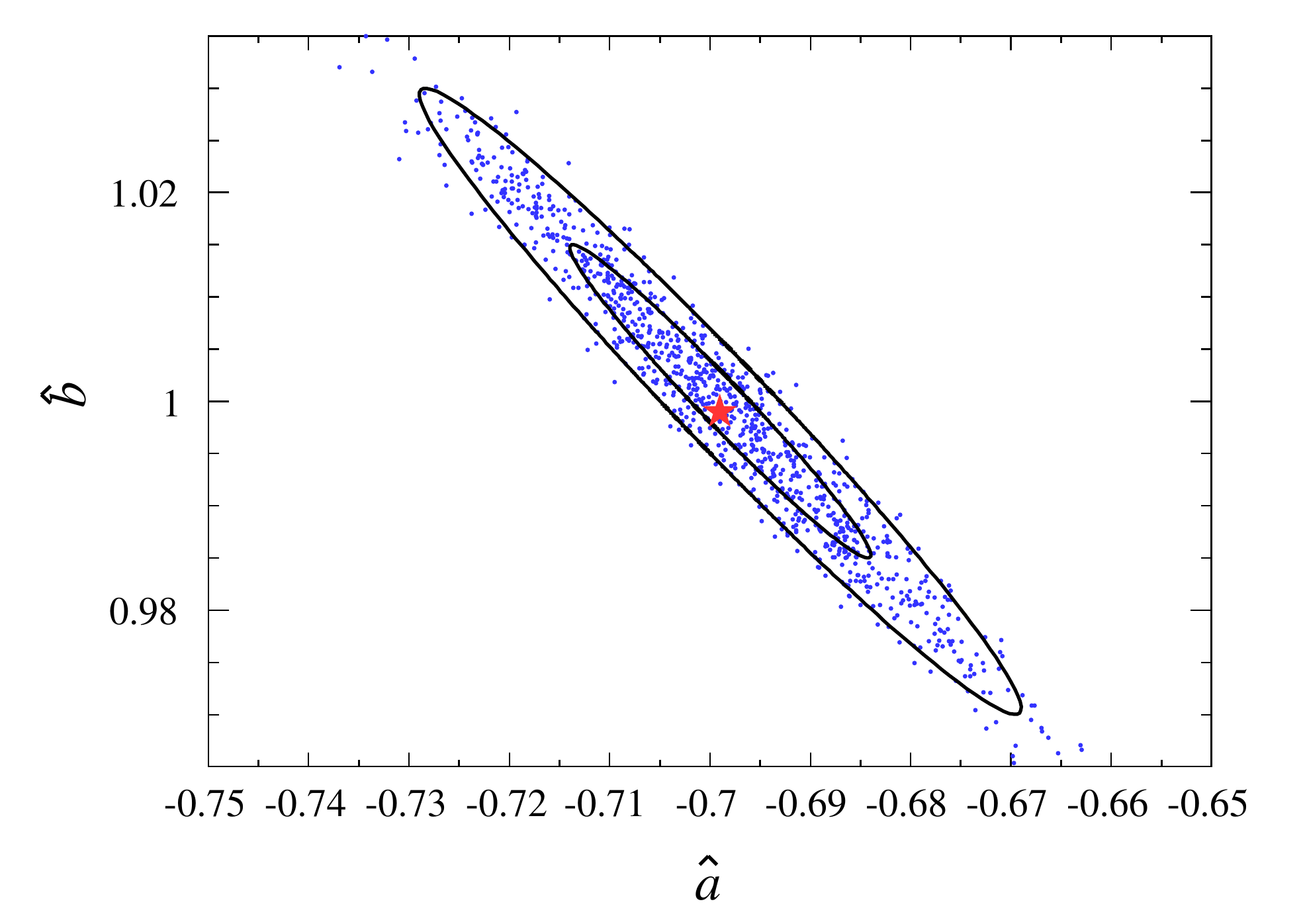}
\caption{\label{fig:lgEcalib} Left: Example of correlation between fluorescence-based energy estimate and shower size estimate, in terms of (decimal) logarithms. Right: Distribution of reconstructed parameters (blue) that establish the linear relationship between $\lg E$ and \lgsref. The red star indicates the average reconstructed value and the black lines are the $1\sigma$ and $2\sigma$ contours.}
\end{figure}
The change of parameters proposed in this study allows for keeping a correct meaning of the covariance matrix for any event topology. 
The use of \lgsref instead of $\sref$ may thus be advantageous for several high-level analyses requiring an accurate event-by-event uncertainty.
The price to pay is to redesign the analyses in terms of the variable \lgsref.
Among other emblematic analyses, we choose to exemplify below such a redesign for the conversion of $\sref$ into energy.
The general strategy currently employed at both the Pierre Auger Observatory and the Telescope Array to carry out this conversion consists in using a set of data with which there are reconstructed shower sizes and an energy assignment by another means.
For the Pierre Auger Observatory, this is done using a special set of air showers that can be reconstructed independently by the fluorescence technique and by the surface detector array~\cite{PierreAuger:2020qqz} while for the Telescope Array, Monte Carlo simulations are used instead.
As an example of how the energy conversion can be derived using the estimated uncertainties, we use the former procedure.
An empirical relationship between the energy measurements from the fluorescence technique and shower size measurements then allows for setting a nearly-calorimetric energy scale.
To derive this relationship, we adapt below the likelihood-based strategy from~\cite{Dembinski:2015wqa,PierreAuger:2020qqz} to a setup that, although simplified, mimics the main features of the Pierre Auger Observatory.

$E$ and $\sref$ are generally related through a power-low relationship, $E=A\sref^B$.
The aim of the procedure is to infer the $a=\lg{A}$ and $b=B$ parameters governing the linear relationship between $\lg E$ and \lgsref:
\begin{equation}
\label{eqn:lgElgS}
    \lg (E/{\rm eV})=a+b\lg\sref.
\end{equation}
The choices $a=-0.7$ and $b=1$ are fairly representative of those found in the Auger data~\cite{PierreAuger:2020qqz}.
To build the relevant likelihood function, we model the distribution of 
$\lg\hat{E}$ and $\lg\hat{S}$ values for events detected simultaneously by both techniques as
\begin{equation}
    \label{eq:raw_event_rate}
    \frac{\dif^2 N}{\dif\lg\hat{E}~\dif\lg\hat{S}}=\iint\dif\lg E~\dif\lg S~R_{\mathrm{FD}}(\lg\hat{E};E)~R_{\mathrm{SD}}(\lg\hat{S};\lg S)~\delta(\lg S,\lg S(\lg E))~\frac{\dif^2 N}{\dif\lg E~\dif\lg S},
\end{equation}
where quantities with a hat denote estimators and, to render the notations more compact, \sref is denoted as $S$.
In this expression, the underlying event rate, $\dif^2N/\dif\lg E\,\dif\lg S$, is folded into the resolution function of the fluorescence detector, $R_{\mathrm{FD}}$, and that of the surface-array detectors, $R_{\mathrm{SD}}$, both expressed in terms of $\lg E$ and $\lg S$, respectively.
In addition, the Dirac delta function guarantees that the underlying relationship given by~\cref{eqn:lgElgS} is satisfied.
The energy resolution of the fluorescence technique is well described as a Gaussian curve with parameters $\mu_E=E$ and $\sigma_E=0.08E$~\cite{PierreAuger:2020qqz}. 
To express the function $R_{\mathrm{FD}}$ in terms of $\lg\hat{E}$ as the primary variable, we thus proceed with the associated Jacobian transformation.\footnote{That is, $R_{\mathrm{FD}}(\lg\hat{E};E)=10^{\lg\hat{E}}\exp(-(10^{\lg\hat{E}}-\mu_E)^2/2\sigma_E^2)\ln 10/\sqrt{2\pi\sigma_E^2}$.}
On the other hand, $R_{\mathrm{SD}}$ can be considered as Gaussian in terms of $\lg\hat{S}$ (log-normal law in $S$ with parameters $\mu_{\lg S}=\lg S$ and $\sigma_{\lg S}$ extracted from~\cref{sec:polar_coords}/~\cref{fig:sref_err_polar_hex}).
Note that to simplify the example application of non-essential ingredients for demonstrating the robustness of the method, only the contribution to $R_{\mathrm{SD}}$ stemming from the detector sampling fluctuations are considered; that is, shower-to-shower fluctuations of $\sref$ are here ignored.
Carrying out the integration in $\lg S$, the probability density function to observe an event with $\lg\hat{E}$ and $\lg\hat{S}$ values is then described as
\begin{equation}
    \label{eq:p}
    p(\lg\hat{E},\lg\hat{S})=\frac{1}{N}\int\dif\lg E~R_{\mathrm{FD}}(\lg\hat{E};E)~R_{\mathrm{SD}}(\lg\hat{S};\lg S(\lg E))~\frac{\dif N}{\dif\lg E},
\end{equation}
where the normalisation factor $N$ is the total number of events detected by the fluorescence technique. 
The likelihood function is finally the product of this expression over the observed $N_{\mathrm{hyb}}$ values of $\lg\hat{E}$ and $\lg\hat{S}$.

To proceed with the likelihood-based strategy, we make use of the bootstrapping technique to substitute the underlying event rate for the observed energies, $\dif N/\dif\lg E \rightarrow E\sum_i\delta(E,\hat{E}_i)$.
In this way, the final expression of the LLH to maximize reads as 
\begin{equation}
    \label{eq:llhcalib}
    \ln \mathcal{L}=\sum_{k=1}^{N_{\mathrm{hyb}}}\ln \left(\frac{1}{2\pi N}\sum_{i=1}^N \frac{1}{\sigma_{E_i}\sigma_{\lg S_i}}\frac{10^{\lg\hat{E}_k}}{10^{\lg\hat{S}_k}}
    \exp{\left[-\frac{\left(10^{\lg\hat{E}_k}-\mu_{E_i}\right)^2}{2\sigma_{E_i}^2}\right]}
    \exp{\left[-\frac{\left(10^{\lg\hat{S}_k}-\mu_{\lg S_i}\right)^2}{2\sigma_{\lg S_i}^2}\right]}
    \right),
\end{equation}
using the notation $\mu_{E_i}=\lg \hat{E}_i$, $\mu_{\lg S_i}=(\lg\hat{E}_i-\hat{a})/\hat{b}$, and where both uncertainty-related terms $\sigma_{E_i}$ and $\sigma_{\lg S_i}$ are estimated by the event-by-event uncertainties.  

We illustrate in~\cref{fig:lgEcalib} the statistical performances of the minimisation of $-2\ln\mathcal{L}$ to recover $a$ and $b$ by generating 1,000 mock samples of events above 1\,EeV until $N_\mathrm{hyb}=3\,000$ events -- typical of the number of hybrid events recorded at the Pierre Auger Observatory -- are drawn above 3\,EeV. 
We use the energy spectrum that is reported in ref.~\cite{PierreAuger:2020qqz}.
The correlation recovered between $\lg S$ and $\lg E$ is shown as the red line in the left panel for one of the mock samples, while the couple of $\hat{a}$ and $\hat{b}$ values recovered for the 1,000 realisations are shown as the blue points in the right panel. 
The red star is observed to be consistent with the injected $a$ and $b$ values while the expected 68\% and 95\% elliptic contours are obtained from the average covariance matrix returned by MINUIT. 
The coverage probabilities from the 1,000 samples are 68.5\% and 97.0\%, consistent with the expectations. The uncertainties in $10^{\hat{a}}$ and $\hat{b}$ are of the same magnitude to those in $\hat{A}$ and $\hat{B}$.

\section{Conclusion}
\label{sec:conclusion}

In this paper, we investigated the accuracy of the uncertainty in air-shower size estimation.
While the uncertainty may be interpreted in the usual way when the errors in the impact point are small compared to the distances from the shower axis to the detectors, this is no longer the case for event topologies where the shower axis is close to a detector.
We showed that in such cases, the LLH-space is highly curved in the ground coordinate system.
For any algorithm that estimates uncertainties using the Hessian matrix formalism i.e., assumes a parabolic shape in the LLH-space, this presents a challenge.
Using a set of toy Monte Carlo air-shower simulations, we showed that this effect is present for both triangular and square array layouts.

However, this effect can be mitigated by choosing a better set of coordinates to describe the shower impact point.
We propose the log-polar coordinate system which is centered on the detector with the largest signal and thus the one around which the LLH-space would be wrapped.
In this coordinate system, the covariance between the shower size and the polar angle coordinate, $\Psi$, is almost decoupled and the LLH-space is better approximated by parabolic curvature near the minimum.
We note that while this will not result in a systematically better estimation of the impact point (nor a worse one), it does produce a better estimation of the shower-size uncertainty.

As this bias is an effect that occurs when the shower axis close to one particular station, we also studied the effects of saturation.
The Cartesian coordinate system produces acceptable estimations of the uncertainty of the air-shower size when the saturated detectors are either ignored or simply accounted for in the LLH calculation.
This comes at the cost of a degraded statistical precision on the air-shower energy and affects the highest-energy events more readily.
When performing a signal recovery on the saturated detectors, the log-polar system is again required and the statistical estimation of the air-shower size (as well as the cosmic ray energy) is improved.
With more modern methods, such as machine learning, or when simply using better hardware with a larger dynamic range, saturation effects can be mitigated or removed entirely.
In any of the cases above, reconstructing using the log-polar coordinate system will produce robust results.


An accurate event-by-event uncertainty of the shower-size estimator may be beneficial for a large variety of high-level analyses.
As an application example, we have shown for instance that the event-by-event uncertainty can be used for the energy calibration of the shower size.
While this application is emblematic for the use of the shower size, any high-level analysis making use of 
event-by-event estimations of uncertainty will benefit from the technique described here.
This includes those which make an event selection based on identifying high-quality events via their estimated uncertainty.
All that is required is to design the analyses in terms of $\lg S$ as the primary variable.

\appendix
\section{Invariance with zenith angle}
\begin{figure}[tbp]
\centering 
\includegraphics[height=7.1cm]{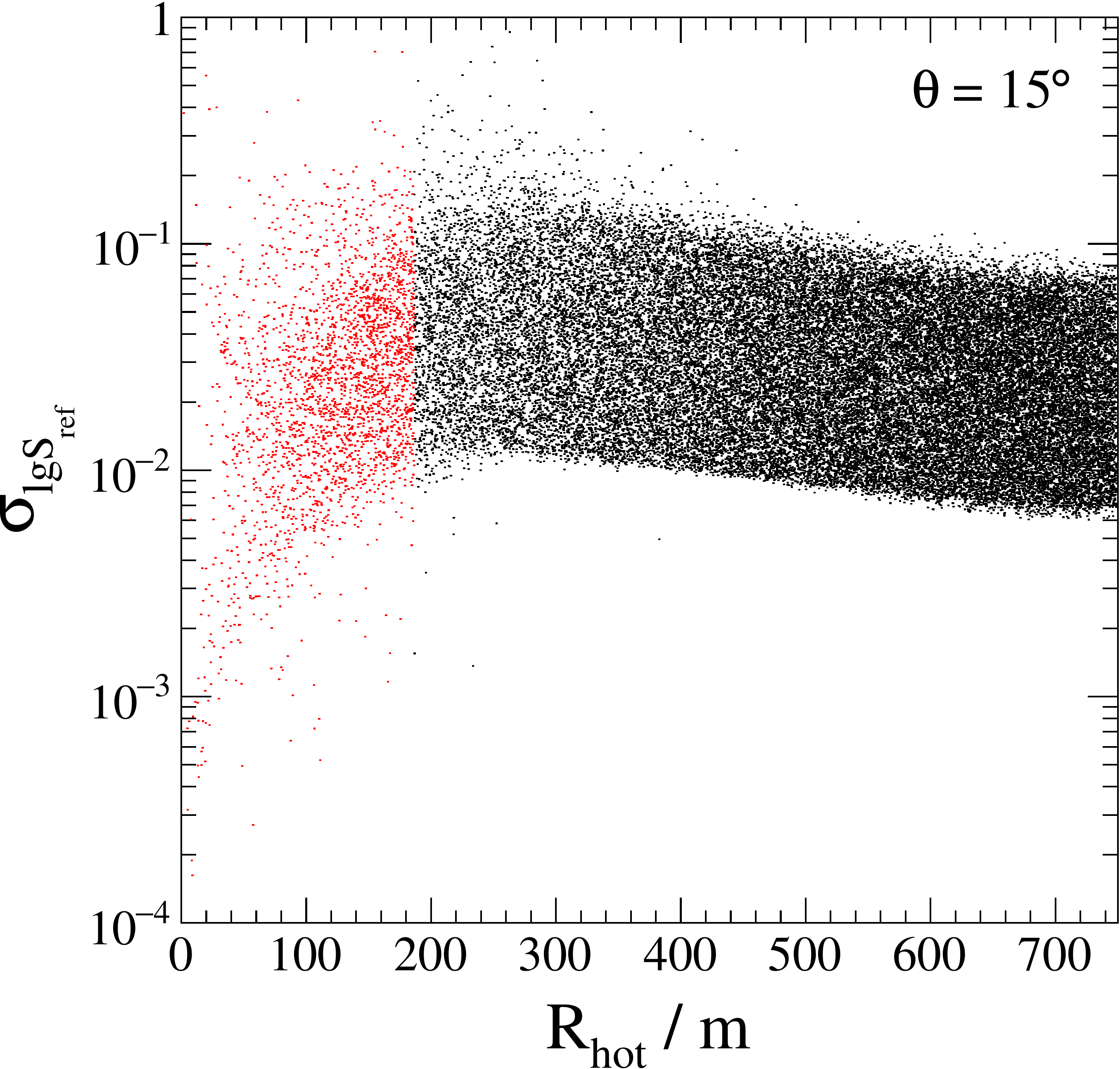}
\hfill
\includegraphics[height=7.1cm]{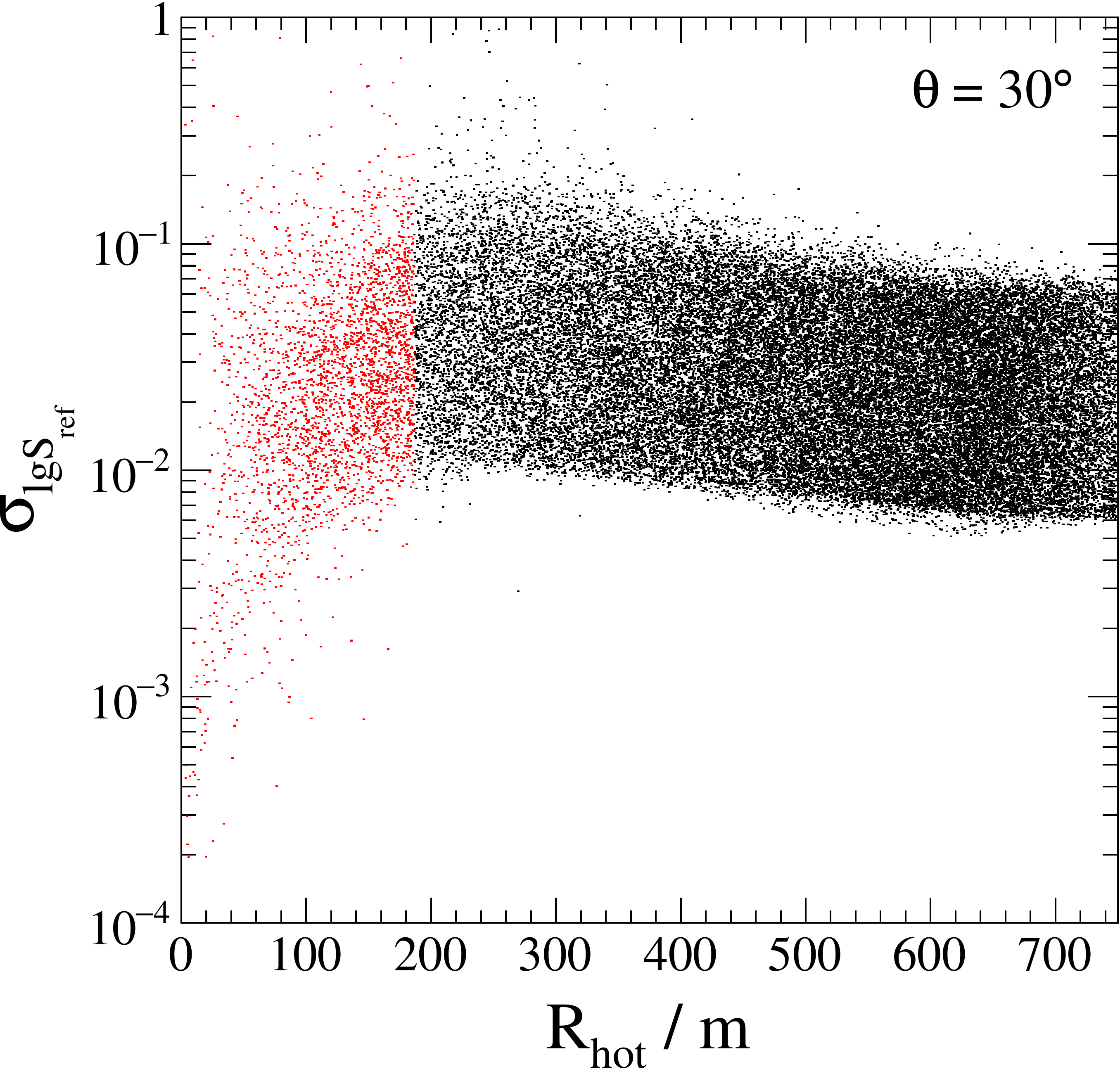}\\
\vspace{0.5cm}
\includegraphics[height=7.1cm]{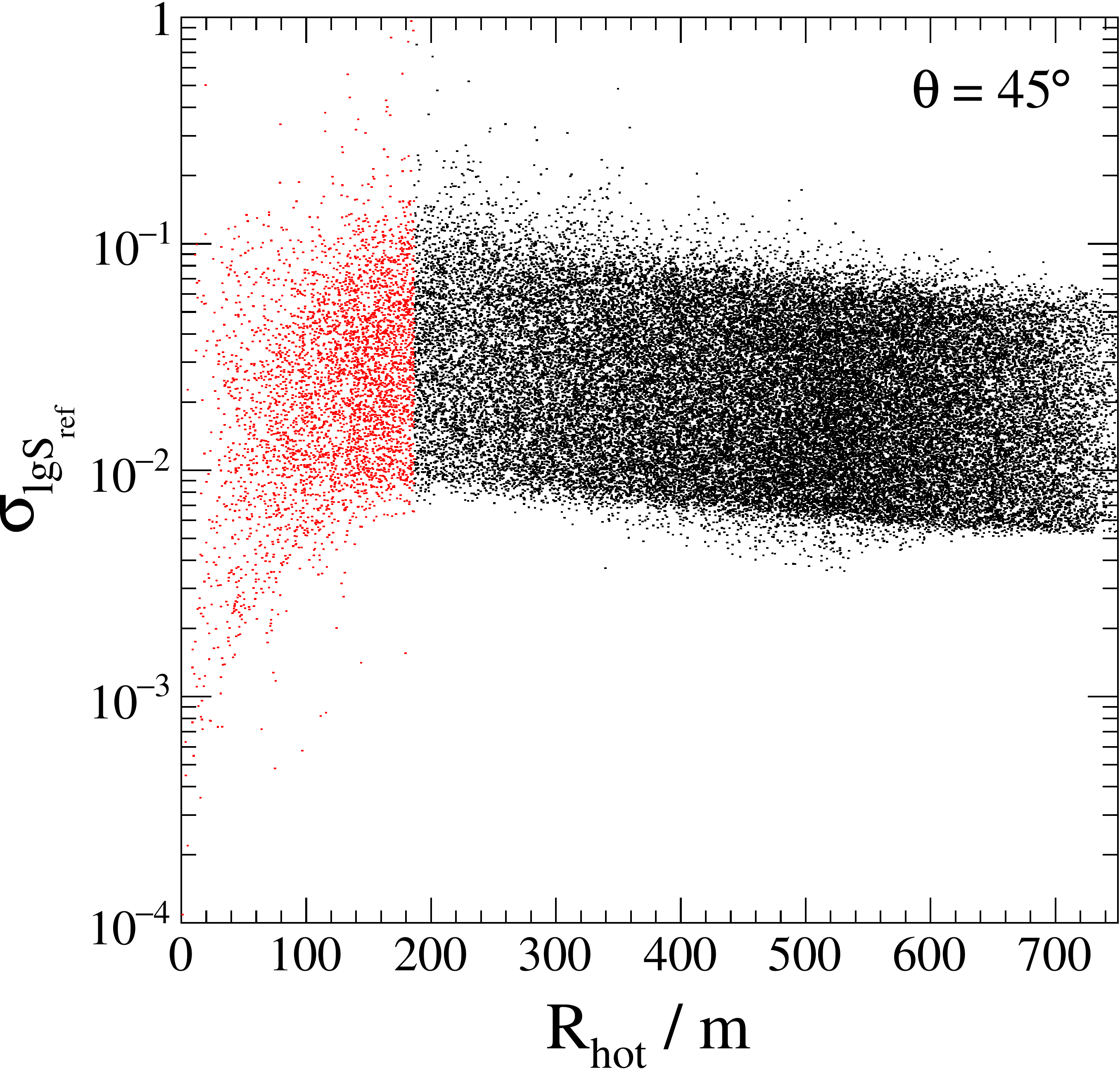}
\hfill
\includegraphics[height=7.1cm]{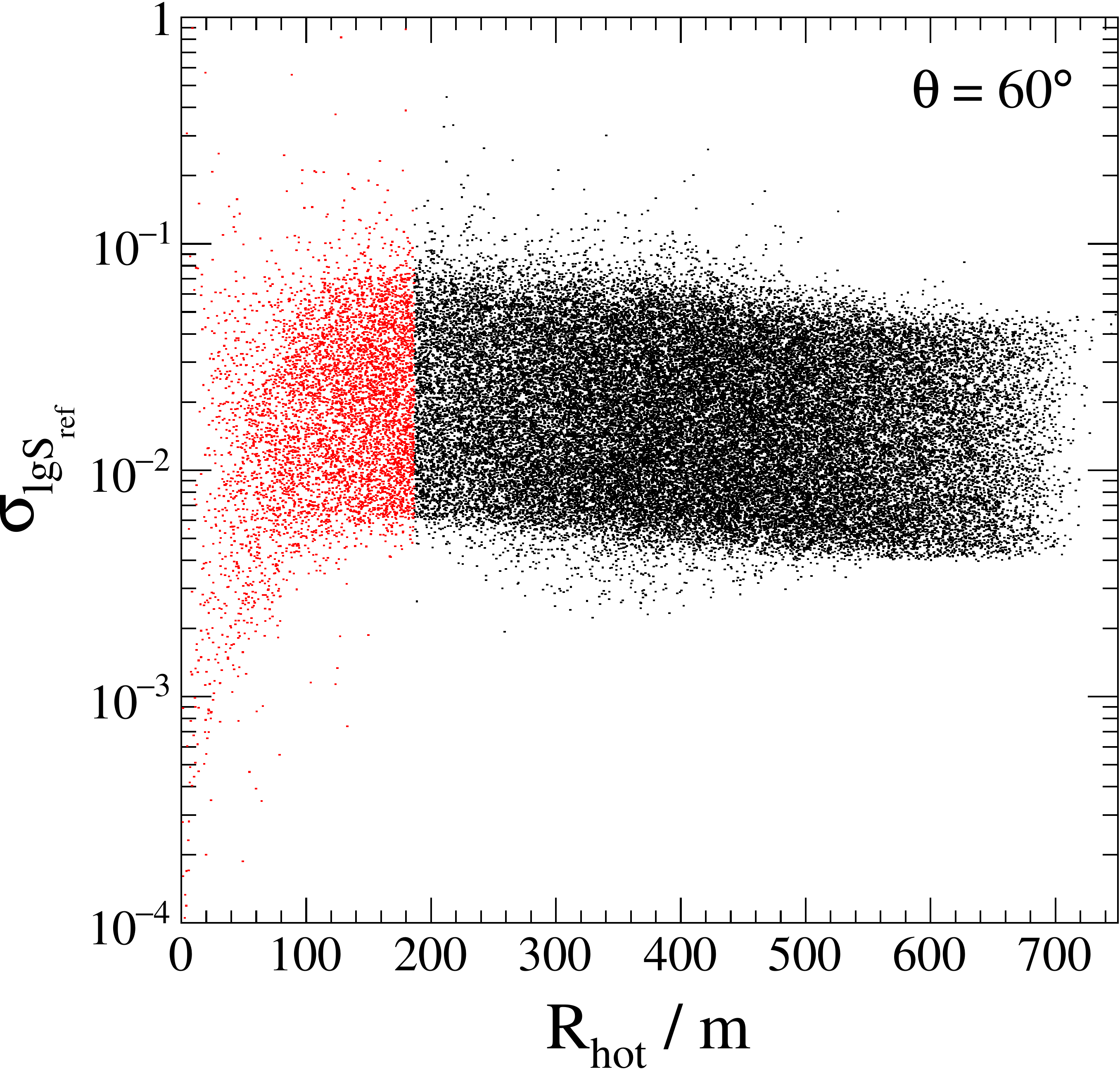}
\caption{\label{fig:zen_invariance} The shower-size uncertainty is shown above for four zenith angles. The red entries highlight the events with $\rhot < 200$\,m. The plots shown here are equivalent to the $40^\circ$ study described in \cref{fig:sref_err_cart_hex}.}
\end{figure}
The underestimation of the shower-size uncertainty was studied above for a fixed zenith angle of $40^\circ$.
However, this effect is present for both more- and less-inclined air showers.
In \cref{fig:zen_invariance}, we show the bias in the relative shower-size uncertainty versus \rhot for four additional zenith angles.
The bias for small values of \rhot is obvious for all of these zenith angles.
Note that the distributions are compressed in the radial direction since, at higher zenith angles, the projected spacing between the detectors in the direction of shower propagation shrinks as $\cos\theta$.
Additionally, this projection effect also leads to more detectors close to the shower axis and thus having a larger signal with smaller uncertainties.
The additional triggered stations that results from this allows for a more accurate estimation of the shower size which can bee seen by the smaller $\sigma_{\lgsref}$ at higher zenith angles.
While we note that the $\beta$ and $\gamma$ parameters of the \LDF may vary with zenith angle and energy (see e.g.~\cite{PierreAuger:2020yab}), we neglected this here to highlight only the effect of changing the zenith angle.



\bibliographystyle{unsrt}
\bibliography{refs}


\end{document}